\documentclass[letterpaper,twocolumn,10pt]{article}
\usepackage{usenix}

\usepackage{tikz}
\usepackage{amsmath}
\usepackage{amssymb} 
\usepackage{booktabs}
\usepackage{multirow}
\usepackage{multicol}
\usepackage{placeins}
\usepackage{tabularx}
\usepackage{hyperref}
\usepackage{xurl}
\usepackage{longtable}

\usepackage{ragged2e} 
\usepackage[flushleft]{threeparttable}
\usepackage{enumitem}
\usepackage{array}
\usepackage[framemethod=tikz]{mdframed}

\newmdenv[
  topline=false,
  bottomline=false,
  rightline=false,
  linewidth=3pt,
  linecolor=gray!40, 
  innerleftmargin=5pt,
  innerrightmargin=0pt,
  innertopmargin=0pt,
  innerbottommargin=0pt,
  leftmargin=5pt
]{keytakeaways}

\begin{document}

\date{}

\title{\Large \bf “It’s like a pet… but my pet doesn’t collect data about me”: Multi-person Households’ Privacy Design Preferences for Household Robots}

\author{
Jennica Li$^{1}$ \quad
Shirley Zhang$^{1}$ \quad
Dakota Sullivan$^{1}$ \\
Bengisu Cagiltay$^{2}$ \quad
Heather Kirkorian$^{1}$ \quad
Bilge Mutlu$^{1}$ \quad
Kassem Fawaz$^{1}$ \\
\\
$^{1}$University of Wisconsin–Madison \\
$^{2}$Koç University Istanbul\\
\texttt{\{hzhang664, dsullivan8, jennica.li, kirkorian, kfawaz\}@wisc.edu} \\
\texttt{bilge@cs.wisc.edu} \\
\texttt{bcagiltay@ku.edu.tr}
}

\maketitle

\begin{abstract}
Household robots boasting mobility, more sophisticated sensors, and powerful processing models have become increasingly prevalent in the commercial market. However, these features may expose users to unwanted privacy risks, including unsolicited data collection and unauthorized data sharing. While security and privacy researchers thus far have explored people’s privacy concerns around household robots, literature investigating people’s preferred privacy designs and mitigation strategies is still limited. Additionally, the existing literature has not yet accounted for multi-user perspectives on privacy design and household robots. We aimed to fill this gap by conducting in-person participatory design sessions with 15 households to explore how they would design a privacy-aware household robot based on their concerns and expectations. We found that participants did not trust that robots, or their respective manufacturers, would respect the data privacy of household members or operate in a multi-user ecosystem without jeopardizing users’ personal data. Based on these concerns, they generated designs that gave them authority over their data, contained accessible controls and notification systems, and could be customized and tailored to suit the needs and preferences of each user over time. We synthesize our findings into actionable design recommendations for robot manufacturers and developers. 

\end{abstract}

\section{Introduction}
The commercialization of smart household robots is no longer a hypothetical concept, but a burgeoning reality. Recent robots such as NEO~\cite{1xNeo2025} and Figure 03~\cite{figure03News2025}, which are advertised as `home robots' and `designed for the home' and promise to handle a wide range of chores and tasks, are becoming more commonplace in the lives and homes of everyday consumers. 

While there is an aggressive push for consumers to integrate these robots into their living spaces,  
it is questionable whether people accept them into their households~\cite{bell2025always}. Prior research has studied privacy maintenance around smart home technologies~\cite{abdi2019more, lau2018alexa, lee2016understanding, lenhart2023you, li2023s}, but these new home-based robots have capabilities beyond smart home devices. Importantly, privacy concerns that users may have about smart home devices may be further intensified in household robots, depending on the types of sensors a robot is equipped with (\textit{e.g.}, cameras or voice recognition)~\cite{lutz2021privacy, windl2024privacy} and the information models it uses to process data (\textit{e.g.}, LLMs, VLMS)~\cite{sullivan2025benchmarking, hu2025information}. 
Furthermore, many household robots are mobile and can navigate spaces unsupervised. Unlike stationary smart devices constrained by fixed sensors, mobile robots can move into or near private areas where users may not want to be seen or heard~\cite{denning2009spotlight, windl2024privacy}.
Given these challenges, it is important to design household robots with a privacy-aware and user-centric approach to foster greater trust and adoption.

There is an increasing, but limited body of research on how people would like their privacy concerns about household robots to be addressed. Thus far, the existing literature explored people’s beliefs about their privacy around household robots~\cite{bacser2024yes, levinson2024surveying, lutz2019privacy}, and a few works have presented what privacy designs and features consumers may want~\cite{windl2024privacy, bell2025always}. However, only a few published studies focused on users’ privacy design preferences for household robots from a multi-user perspective. Prior research on the security and privacy of smart home devices shared among users in the same home suggests that users express varying levels of privacy concerns and disagree about how to control the device~\cite{geeng2019s}. Additionally, to our knowledge, there is no existing work on privacy-aware designs for household robots that uses a co-design method, where participants interact with a robot in person as part of the design process, despite this being a common practice in the human-robot interaction field.  

In this work, we address how to design privacy-aware household robots for multi-user environments. To do so, we pose three research questions: 

\textbf{RQ1}: What factors shape households' perceptions of the privacy of a shared household robot?

\textbf{RQ2}: What designs do households prefer in privacy-aware shared household robots?

\textbf{RQ3}: What mechanisms affect how households might implement privacy-aware options?

Toward this end, we conducted in-person co-design sessions with 15 households (\textit{i.e.}, two or more participants who resided in the same space).  The study took place in a lab space modeled after a living room, where participants interacted with a physical robot. We found that households were concerned about allowing a household robot, and their respective manufacturers, to collect and use their data without restrictions. Participants also felt uncertain about a robot's ability to function in a multi-user setting, where different users might have different preferences or levels of comfort. To mitigate these concerns, participants suggested a variety of privacy designs that would uphold individual users' authority and awareness over the robot's actions and data collection practices. We summarize the design recommendations from participants into the following:

\noindent
\textbf{User Profiles}: The robot should offer users the option to build personal, customizable profiles that can only be accessed through authentication.

\noindent
\textbf{Mapping}: The robot should allow users to customize their robot's physical access to match household schedules, thus restricting access to areas of the home during specific times of day. Users can also define ``safe-zones'' for robots when they are not performing tasks, or pre-defined paths for robots to follow.

\noindent
\textbf{Data Storage}: The robot should only allow authorized and authenticated users to access stored data that is relevant to them (\textit{e.g.}, if they appear in the captured video, if they requested the robot to save some information). The data retention period should be set by the user based on their preferences. Allow users to only store their data locally. 

\noindent
\textbf{Data Collection}: The robot should present accessible overviews of its data collection practices. In addition, the robot should prioritize low-fidelity sensor data (\textit{e.g.}, motion sensors) over content-rich visual or audio data for routine operation, and require explicit wake words to activate cameras. To address users' concerns, provide physical cover and a clear visual/audio signal when sensors are on. 

\noindent
\textbf{Notifications}: The robot should be able to understand when, what, and who to send notifications. Only relevant users should be notified, and in some cases, the robot should deliver notifications to a connected device (\textit{e.g.}, phone).

Although households highlighted different privacy designs that could be valuable in a shared household robot, this did not mean that they would implement these designs in a household robot they owned. Their ideas around design adoption were influenced by four key mechanisms: 1) whether the robot's benefits were enough to outweigh loss of privacy or the extra effort needed to maintain privacy; 2) participants' social treatment and attachment to the robot; 3) conflicts around how to manage the robot amongst household members; and 4) the presumed effects of long-term exposure and experience with the robot on privacy preferences.

\section{Related Work}



\subsection{Privacy Perception of Smart Home Device}
Over the past decade, smart home devices have become increasingly commonplace in U.S. households~\cite{horowitzSmartDevices2025}. From smart speakers to Bluetooth-enabled appliances, these devices offer small conveniences that can be integrated into a household’s routines. However, introducing smart home devices into shared living spaces exposes all household members or guests to potential privacy risks, including undisclosed passive data collection~\cite{lee2016understanding, tabassum2019don, yao2019defending, abdi2019more, malkin2019privacy, zheng2018user, lau2018alexa, zeng2017end, li2023s, apthorpe2018discovering,bolton2021security,seymour2023legal,khezresmaeilzadeh2024echoes} and improper use of collected data~\cite{malkin2019privacy, hudig2025rights, le2024towards, acosta2022survey}. Even if only one household member is a primary user of the device, non-users may still be exposed to risks simply by being near the device’s data sensors~\cite{bernd2020bystanders, ahmad2020tangible, marky2020don, mare2020smart, pattnaik2024security, saqib2025bystander, yao2019privacy, alshehri2022exploring}. Prior work also shows that these perceptions affect customers' device purchase behaviors~\cite{emami2019exploring, emami2023consumers, barbosa2020privacy}. 
Building on these perceptions, prior works propose various directions to explore how households discuss and manage privacy conflicts~\cite{zeng2019understanding, park2023nobody, lenhart2023you}. Researchers also conduct different systems to help users better evaluate and manage their privacy when using smart home devices~\cite{wang2025privacyguard, aaraj2024vbit, albayaydh2024co}, or introduce the privacy labels for both users and developers to better understand smart home devices~\cite{emami2020ask}. Prior research on the security and privacy of shared smart home devices within the same home suggests that users express varying levels of privacy concern and disagree on how to control the devices~\cite{geeng2019s}. Our study aims to extend this work to household robots in shared spaces by investigating how households might design robots to uphold their privacy. 

\subsection{Security and Privacy of Household Robots}
Compared with existing smart home devices, household robots may offer unique affordances that enhance their versatility as day-to-day tools. Whereas smart home devices can sometimes collect data from a nearby distance, they often cannot move unless a user picks them up and relocates them. Household robots, by contrast, combine mobility with increasingly human-like interaction capabilities enabled by GenAI (\textit{e.g.}, large language models, vision-language models)~\cite{zitkovich2023rt, driess2023palm, singh2022progprompt, wake2023chatgpt}, allowing them to proactively move through and engage across domestic spaces. When coupled with programs that mimic more human-like socialization and learning, household robots offer a wide variety of use cases for various populations, including providing physical and social support to the elderly~\cite{wada2007living, shibata2011robot, moyle2018care, carros2020exploring} and offering tutoring or social practice to children~\cite{han2005educational, randall2019survey, kennedy2015robot, kennedy2016social, belpaeme2018social}.

Importantly, these affordances also exacerbate household privacy risks associated with other smart technologies and introduce distinct privacy risks. Mobility may increase the likelihood that a household robot may encounter, collect, and process data or information that was not meant to be overheard or recorded~\cite{denning2009spotlight,windl2024privacy}. They may also become more accessible to household members who do not want to interact with them~\cite{lutz2021privacy}, or should have restricted access to them (\textit{e.g.}, young children~\cite{levinson2024surveying}). Regarding the implementation of advanced GenAI software, scholars have conducted multiple comprehensive analyses of its vulnerabilities~\cite{shi2024large}. Some of these concerns center on the possibility of models being attacked ~\cite{carlini2021extracting, staab2023beyond} or generating responses that contain incorrect or potentially harmful information~\cite{huang2025survey}. Other researchers have raised concerns about the amount of data needed to train a GenAI model effectively; if certain users in the household are unwilling to allow their robot access to their data, this could result in a less productive or useful robot than expected~\cite{stapels2023never}. Expecting that users might be lenient about granting a robot access permission to their data if it improves performance could be unsound, as both Levinson et al.~\cite{levinson2024snitches} and Bell et al.~\cite{bell2025always} found that people’s privacy concerns around robots intensified when asked to imagine the robot within the context of a home environment.

While household robots may be a recent development, the privacy concerns people and households have about them are not well-founded~\cite{denning2009spotlight, collins2024socially, guerrero2017cybersecurity, pagallo2018rise, rueben2018themes, levinson2024snitches, levinson2024surveying, lutz2019privacy, bacser2024yes, tang2022confidant}. Mobile vacuum robots, which tend not to boast a wide range of features beyond what is needed to accomplish their cleaning, have already been the center of multiple privacy-breaching incidents~\cite{futurismRobotVacuum2025, techreviewRoomba2022}. Additionally, manufacturers have not shown that they have addressed users’ existing concerns in a meaningful way; in their analysis of eight robotics companies’ terms of use, Chatzimichali et al.~\cite{chatzimichali2020toward} identified inconsistencies in how robots’ privacy policies were communicated to users. Thus, more work is needed to support the creation of privacy-aware household robots that households would feel comfortable with and confident in.

\subsection{The Current Study: Improving the Privacy and Security of Household Robots}
Although there is extensive literature on users’ privacy and security concerns, few studies have examined designs that can address or mitigate these concerns in household robots. In the literature on smart home devices, multiple researchers have invited participants and households alike to ideate and co-design privacy features they would like to implement~\cite{yao2019defending, yao2019privacy, abdi2019more, marky2024decide}. Yet, little work of this nature exists for household robots. One exception is Bell et al. ~\cite{bell2025always}, who found that participants were concerned about the robots' autonomous interactions and wanted greater control over them. Notably, there is also a dearth in literature that explores privacy designs from a multi-user perspective. Although Bell et al. presented scenarios in their research involving more than one user, the designs were generated through one-on-one online interviews. This study aims to take an in-person co-design approach to exploring multi-user perspectives on designing privacy-aware household robots. By co-designing with a household robot, participants can better understand which designs are realistic by seeing their ideas in action~\cite{stegner2023situated, ostrowski2021long}. Additionally, this study focuses on households, represented by two or more participants living in the same space, to account for privacy concerns, preferences, and negotiations around household robots that may be unique to multi-person use or environments.  
\begin{figure*}[!ht]
    \centering
    \includegraphics[width=\linewidth]{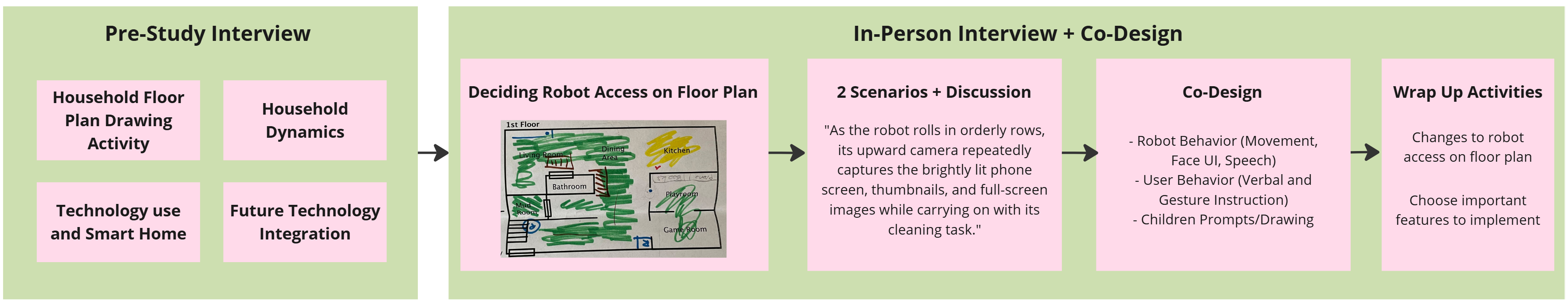}
    \caption{Overview of Study Procedure. Participants first completed a pre-study interview covering their household management and their technology perception. They are then invited to an in-person interview with a co-design session.}
    \label{fig:method}
\end{figure*}

\section{Methods}
We conducted an in-person co-design study to understand people's preferences for a privacy-aware social robot in households. Before attending the study, participants completed an online pre-study interview, where they drew their household floor plans and discussed their use of smart technologies.  Figure~\ref{fig:method} presents an overview of our study.

\subsection{Recruitment and Participants}
The research team recruited participants via research e-mail listservs and physical fliers posted in community spaces (\textit{e.g.}, community centers). Participants had to be (1) living in the same residence with at least one other person for at least two days each week, (2) residing in the U.S., (3) fluent in English, and (4) at least 18 years old (unless they were minor children of participating parents). Only one person from each household completed the pre-study. Participants completed their pre-study interviews between May and July 2025 and their in-person co-design sessions, along with at least one other household member, between July and August 2025.


The in-person study consisted of 36 participants (55.6\% women/girls, 44.4\% men/boys) across 15 households, with each household represented by two to four people. Eight households included participants under 18 years of age; all minors were the children of the adult participants. Most participants identified as White (75\%), followed by Asian or Pacific Islander (11.1\%), or multiracial (5.6\%). On average, participants were 32 years old, ranging from 3 to 67. 14 of the households (93.3\%) included at least one adult participant with a four-year degree or higher. 7 (46.7\%) reported an annual income of \$79,999 or less. 

\subsection{Materials and Procedure}
We employed co-design methods popularized in the HRI community~\cite{antony2023co, claudio2017interaction, obaid2023collective, yao2019defending}, allowing participants to interact with and ideate on the technologies they are designing. The following sections describe our study materials, procedures, and the household robot used in the co-design process.

\subsubsection{Co-Design Materials}
Prior to each in-person study session, the research team prepared a set of materials to support participants’ design idea generation, including simple art supplies (\textit{e.g.}, plain paper, printable with various visuals that the robot could display, markers), a miniature, 3-D printed version of the study robot, and a printed floor plan that one of the household's participants had self-drafted during their online pre-study interview. 

\begin{table}[h!t]

\centering
\scriptsize
\begin{threeparttable}
\caption{Household and Participant Information}
\label{demo}
\newcolumntype{P}[1]{>{\raggedright\arraybackslash}p{#1}}

\begin{tabularx}{\columnwidth}{@{}P{0.2cm} P{0.25cm} P{0.5cm} P{0.2cm} P{0.5cm} P{1.7cm} P{1.2cm} P{0.95cm}@{}}

\toprule
\textbf{\shortstack{HID}} & 
\textbf{PID} & 
\textbf{Rel.} & 
\textbf{Age} & 
\textbf{Gender} & 
\textbf{Race} & 
\textbf{Edu.} & 
\textbf{Income} \\
\midrule
\multirow{2}{*}{1} & 1\_1 & n/a & 32 & Woman & White & Doctoral & 80-89k\\
& 1\_2 & Partner & 36 & Man & Other & Doctoral & \\ 
\midrule
\multirow{2}{*}{2} & 2\_1 & n/a & 31 & Woman & White & Master's & 60-69k \\
& 2\_2 & Partner & 29 & Man & prefer not to say & Master's & \\ 
\midrule
\multirow{2}{*}{3} & 3\_1 & n/a & 50 & Man & White & Master's & 100-149k\\
& 3\_2 & Child & 21 & Man & White & Some college & \\ 
\midrule
\multirow{3}{*}{4} & 4\_1 & n/a & 44 & Woman & White & Master's & 150k+\\
& 4\_2 & Partner & 47 & Man & White & Bachelor's & \\
& 4\_C & Child & 14 & Girl & White & n/a & \\ 
\midrule
\multirow{4}{*}{5} & 5\_1 & n/a & 38 & Woman & White & Professional & 70-79k \\
& 5\_C1 & Child & 8 & Boy & White & n/a & \\
& 5\_C2 & Child & 11 & Girl & White & n/a & \\
& 5\_C3 & Child & 12 & Girl & White & n/a & \\ 
\midrule
\multirow{3}{*}{6} & 6\_1 & n/a & 44 & Woman & White & Bachelor's & 150k+\\
& 6\_2 & Partner & 43 & Man & White & Bachelor's & \\
& 6\_C & Child & 15 & Girl & White & n/a & \\ 
\midrule
\multirow{3}{*}{7} & 7\_1 & n/a & 44 & Woman & White & Doctoral & 70-79k\\
& 7\_2 & Partner & 41 & Man & Asian or Pacific Islander & Bachelor's & \\
& 7\_C & Child & 3 & Boy & Multiple\textsubscript{1} & n/a & \\ 
\midrule
\multirow{2}{*}{8} & 8\_1 & n/a & 64 & Woman & White & Associate's & 100-149k \\
& 8\_2 & Partner & 67 & Man & White & High school & \\ 
\midrule
\multirow{3}{*}{9} & 9\_1 & n/a & 50 & Woman & White & Bachelor's & 150k+ \\
& 9\_2 & Child & 19 & Man & White & Some college & \\
& 9\_C & Child & 16 & Boy & White & n/a & \\ 
\midrule
\multirow{2}{*}{10} & 10\_1 & n/a & 32 & Woman & White & Bachelor's & 100-149k \\
& 10\_2 & Partner & 35 & Man & White & Bachelor's & \\ 
\midrule
\multirow{2}{*}{11} & 11\_1 & n/a & 25 & Man & White & Bachelor's & 40-49k \\
& 11\_2 & Partner & 25 & Woman & White & Bachelor's & \\ 
\midrule
\multirow{2}{*}{12} & 12\_1 & n/a & 34 & Man & White & Master's & 40-49k \\
& 12\_2 & Partner & 32 & Woman & White & Bachelor's & \\ 
\midrule
\multirow{2}{*}{13} & 13\_1 & n/a & 58 & Woman & White & Doctoral & 40-49k\\
& 13\_C & Child & 17 & Boy & Black or African American & n/a & \\ 
\midrule
\multirow{2}{*}{14} & 14\_1 & n/a & 46 & Woman & Asian or Pacific Islander & Master's & 100-149k  \\
& 14\_C & Child & 11 & Girl & Multiple\textsubscript{2} & n/a & \\ 
\midrule
\multirow{2}{*}{15} & 15\_1 & n/a & 21 & Woman & Asian or Pacific Islander & Bachelor's & 60-69k \\
& 15\_2 & Sibling & 23 & Woman & Asian or Pacific Islander & Bachelor's & \\
\bottomrule
\end{tabularx}
\begin{tablenotes}
    \item \textsubscript{1} Asian or Pacific Islander and White, \textsubscript{2} Asian or Pacific Islander and African American
\end{tablenotes}
\end{threeparttable}
\end{table}

\subsubsection{Study Procedure}
Two research team members ran the in-person study sessions in a lab space that resembled a living room: one team member facilitated the session, while the other controlled the robot. Every session began with the study facilitator obtaining written consent from adult participants and, when including minors, verbal assent from minor participants. Afterward, participants completed a set of surveys, including demographic questions. Once participants had completed their surveys, they reviewed and corrected the household floor plan the research team created based on the pre-study, ensuring accurate representation of their own households. The research team then introduced the participants to the robot and asked participants to interact with it. The study facilitator then asked participants to use their floor plan to determine which areas of their home they would allow the robot to access. Once the household reached a general agreement on their access mapping, the facilitator asked them about their use and perceptions of the existing technological devices in their household. Then, we introduced participants to at least two hypothetical scenarios involving a household robot: one scenario on data privacy and one on interpersonal privacy, though some households discussed more if time allowed. For households with young children, the facilitators presented simplified versions of scenarios. This is an example of an interpersonal privacy scenario: 

\begin{quote}
Context:
Anna paces in front of her desk at night, dictating tomorrow’s to‑do list out loud. Down the hallway, the household robot is rolling past on its scheduled trash‑bin pickup; its long‑range microphones remain active for voice commands.

Anna (raising her voice to capture the robot's attention): ``\textit{Robot, remind me at 7 AM to email the team the draft and order balloons for Emma's surprise party.}''\\
\end{quote}

After introducing each scenario, the facilitator asked the households for their initial reactions and potential concerns, and then asked them to brainstorm design features to mitigate or address them. Engaging in the co-design process, the facilitator sometimes followed up with specific questions about households' preferences for how the robot should react or respond, probing around how the robot should move or indicate that it was acting in a privacy-aware manner (\textit{e.g.}, visual indications on its screen, verbal cues). While the scenarios were presented to guide participants’ discussions, the facilitator allowed households to ideate and expand on other ideas or issues that arose organically during conversations. 

Finally, the facilitator asked households to review their floor plan again and indicate whether their preferences regarding the robot’s access had changed after interacting with it. Then, the facilitators asked the households to determine their most desired privacy design features and ended the study with several questions about the participants’ experience with the robot. 

Households received \$50 for completing the study, and were compensated an additional \$2-5 dollars for parking if needed. Sessions took about 90 minutes to two hours to complete. Our institution's Institutional Review Board (IRB) approved our study design and materials. 

\subsubsection{Robot Selection}
The research team selected Temi V3~\cite{robotemi_temi_v3_2026} to represent a household robot that participants might encounter, as it could move on its own and map out spaces. 
During the study, one researcher controlled Temi using the Wizard-of-Oz technique\cite{dahlback1993wizard}, meaning all its movements, reactions, and responses were remotely controlled by a researcher in the room. This allowed the team to program Temi to act out certain scenarios or design features (\textit{e.g.}, leaving the room) and manipulate the visuals on the attached touchscreen interface to gauge participants’ reactions. 
The researcher informed participants early on that they controlled Temi and encouraged them to suggest design features for Temi to enact.
Throughout this work, we will refer to Temi as `the robot’ and discuss participants’ design suggestions from a more general perspective. 

\subsection{Data Analysis}
We performed inductive thematic analysis ~\cite{braun2006using} on the qualitative data collected during the study. 
The research team generated transcriptions of the audio recordings using noScribe~\cite{kaixxx_noScribe_2026}, an open-source tool that transcribes audio using a local Whisper model. Each transcription was then manually reviewed and corrected by at least two members of the research team. Once the transcriptions were cleaned, the first author independently coded a randomly selected transcript and developed an initial codebook. Afterwards, the second author independently coded the same transcript using the generated codebook, then met with the first author to discuss and resolve any disagreements. This iterative process continued until the coders had reached consensus on all 15 transcripts.
To identify overarching themes in the data, the first two authors met with the third author to discuss the final codebook.

	To maintain participant confidentiality, we refer to households and participants using codes. Households are numbered 1–15, and participants are labeled with their household number followed by an underscore and an additional identifying number (from 1-4). We identify minor participants with the letter `C'. 
     For instance, the participating adult from Household 5 is referred to as 5\_1, while their three minor children who also participated are referred to as 5\_C1, 5\_C2, and 5\_C3.




\begin{table*}[ht]
    \centering
    \scriptsize 
    \renewcommand{\arraystretch}{1.5} 
    \caption{Summary of All Design Patterns}
    \newcolumntype{P}[1]{>{\raggedright\arraybackslash}p{#1}}

    \label{summary_of_design}
    \begin{tabularx}{\textwidth}{llX}
        \toprule
        \textbf{Thematic Group} & \textbf{Category} & \textbf{Examples of User Suggestions} \\
        \midrule
        \multirow{3}{*}{\parbox{1.5cm}{\raggedright Private Space}} & Area-specific limitations & Ask for permission to enter specific areas; Do not collect data afar for red zone; Hold still at charging location; Predetermined pathways; Only allow recording in specific areas; Turn away from `yellow' areas. \\ \cline{2-3}
        
        & Time-specific limitations & Only function during specific time; Only collect data during specific times; Set `active' period. \\ \cline{2-3}
        
        & Safety spots and homebases & User-requested Relocation; Send away to different location; Stationed in common area; Multiple homebases for convenience. \\ 
        
        \midrule
        \multirow{3}{*}{\parbox{1.5cm}{\raggedright Privacy within and Beyond Family}} & Visual/Audio/Haptic Recognition & Biometric recognition and-or activation; Facial or Voice Recognition; Scan the room for users or situations before functioning; Biometric. \\ \cline{2-3}

        & Protections between users & Profiles for different users; Allow users to set password for accessing information; Notify only relevant user; Notify on another device (do not show on screen); Notify when other user is accessing camera; Vague notifications; Log user access to Temi's data; Only share data with authorized user. \\ \cline{2-3}

                 & Outsider access & Share data only with family; Do not record or interact with non-household members; Guest Mode. \\

        \midrule
        \multirow{11}{*}{\parbox{1.5cm}{User-Initiated Actions}} & User-initiated data collection & Allow users to pause data collection temporarily; Privacy Mode; Offline mode; Only collect data when given explicit permission; Wait for user initiation. \\ \cline{2-3}

        & User-Initiated Training & User-Initiated Training. \\ \cline{2-3}

        & Physical `control' & Physical barrier to cover the camera; Physical button or indicator to pause recording; Physical Buttons; Physical Representation of Storage; Physical Buttons; Power off completely. \\ \cline{2-3}

        & Physical robot reactions & Relocate sensor; Lower the screen. \\ \cline{2-3}

        & User-initiated data management & Offer ability to review, manage, or delete data; Ask user how to manage data; Get permission to delete data. \\ \cline{2-3}

        & Limited camera activation & Default camera off; Do not record; Passive sensor data; Record video but not audio. \\ \cline{2-3}


         & Data retention timing & Immediately delete data; Delete data after a predefined period; Delete data for a time period. \\ \cline{2-3}

        & Reasons for retaining data & Keep data for task intention; Learn or improve from the information; Retain essential data. \\ \cline{2-3}

        & Data sharing limitations & Do not share data; Do not upload data online or to manufacturer without permission; Do not use data to train; Only share or use data when given permission. \\ \cline{2-3}

        & Data storage preferences & Store data locally only; Securely store data. \\ \cline{2-3}

        & Anonymizing data & Anonymize data depends on recipents; Anonymize data. \\
        \midrule

        \multirow{5}{*}{\parbox{1.5cm}{Robot-Initiated Actions}} & Explanations of policies \& practices & Explain data retention policies; Explain purpose; Baseline manual explaining purpose; Explain protocol and functions; Remind user about data management; Be transparent about data collection. \\ \cline{2-3}

        & Data-related notifications & Notify user of data collection; Notify when data collection is not possible. \\ \cline{2-3}

        & Visual notification preferences & Visual indication (camera view, color, facial, icon or symbol, transcript, unspecified). \\ \cline{2-3}

        & Audio notification preferences & Audio indication (non-verbal); Verbal indication; Announce presence. \\ \cline{2-3}

        & Indicators of not recording & Other visual indication of NOT recording; Asleep' or 'blacked out' to indicate NOT recording. \\


        \bottomrule
    \end{tabularx}
\end{table*}
\section{Privacy Designs for Household Robots}
We first explore why and how households might design a shared household robot to be more privacy-aware. We present overarching reasons participants sought privacy designs, then examine the themes their designs fell into. We follow this by discussing how participants' perceptions of the robot influenced whether they would adopt their designs in practice.  
\subsection{Reasons for Privacy Designs}
The following section presents the four most notable motivators for households' suggestions of privacy design features.

\noindent
\textbf{Data Privacy and Transparency.}
Almost all of the design features suggested by participants were, in part, motivated by concerns about a robot’s ability to uphold users’ data privacy and security, and to do so transparently. Participants cited concerns about whether the robot’s various sensors (\textit{e.g.}, camera, passive listening) could lead to unwanted data collection. 
They noted that the robot’s mobility increased its potential to encounter and record private moments that other stationary technology would not have access to. It was important to participants that a household robot not only be equipped with functions to protect users’ privacy, but also have transparent indicators or policies about its actions. 

\noindent
\textbf{Multi-user Concerns.}
Participants worried that a robot shared among multiple users could lead to privacy breaches within the household (\textit{e.g.}, one user gaining access to another user’s private conversations with the robot). Participants also considered potential issues with having a robot that could collect data on or provide information to household guests or other bystanders (\textit{e.g.}, a mail carrier). As a result, participants sought out designs that could protect themselves, others in their household, and people outside of their household that might come into contact with the robot. 

\noindent
\textbf{Household Routines and Structure.}
Participants often situated their design suggestions within the context of their personal and shared household routines and living spaces. This led participants to consider how they might design a robot that could satisfy different household members’ needs and privacy preferences, or to suggest restricting the robot's entry to specific areas of the home under certain circumstances. 

\noindent
\textbf{Distrust in Technology Manufacturers.}
Many participants did not believe that dominant technology companies and manufacturers had their consumers’ privacy and best interests in mind. These beliefs stemmed from personal experiences with other devices (\textit{e.g.}, smartphones, conversational agents) or information acquired from various media outlets or via word-of-mouth. 
As a result, participants desired more restrictive or protective designs that could limit the level of interference or control manufacturers might yield over their devices and data.

\noindent

\subsection{Types of Privacy Design}
\begin{figure*}[!htbp]
    \centering
    \includegraphics[width=0.84\linewidth]{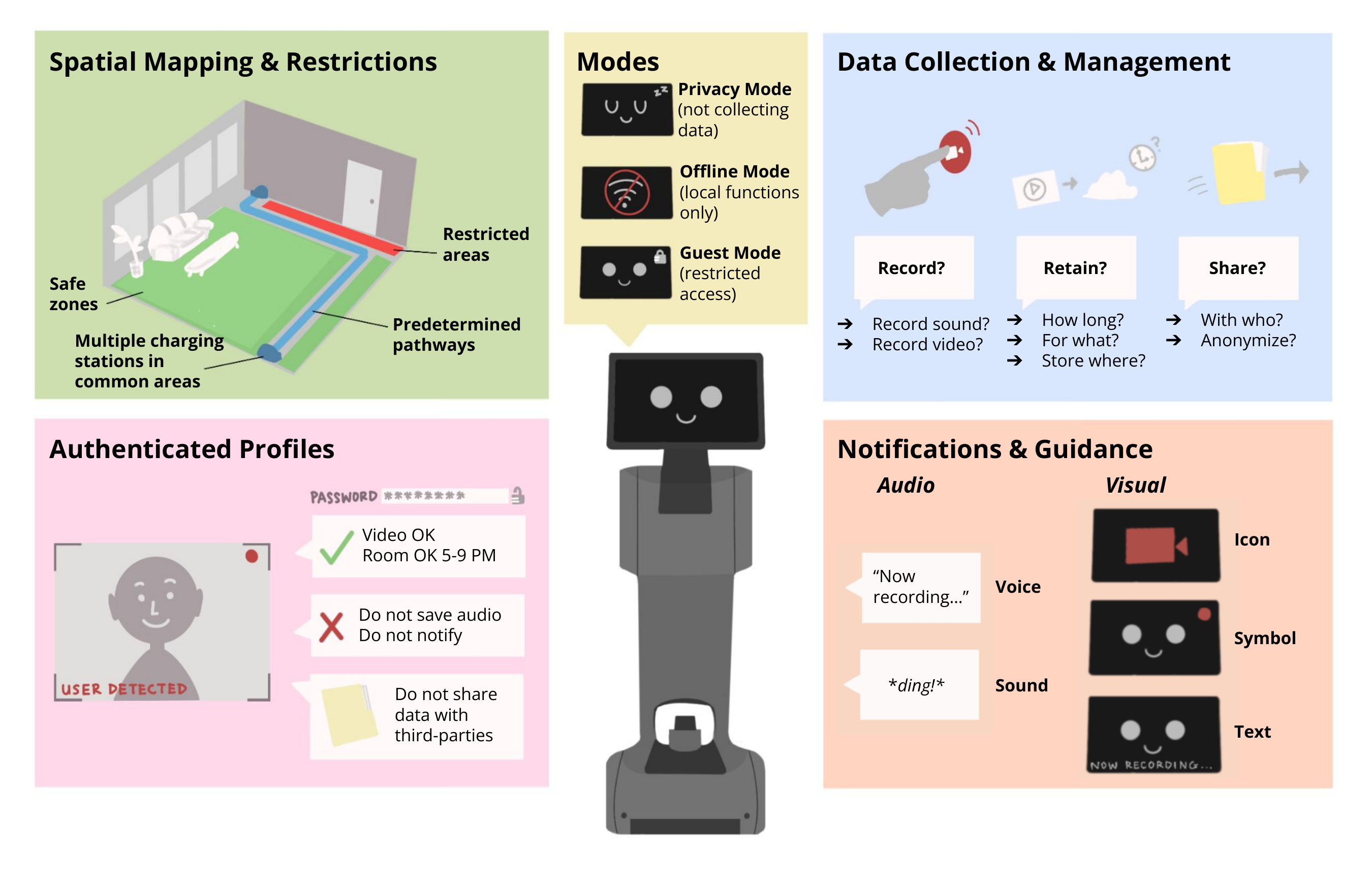}
    \caption{A set of privacy-aware features on a robot, including different modes, a mapping system, a data collection and management system, a user profile and authentication system, and a notification system.}
    \label{fig:temi}
\end{figure*}

We report the privacy-related design themes and features that participants suggested throughout their interviews. We present these themes in two subsections: designs focused on how the robot should operate based on its environment, and designs focused on how the robot should interact with its user. 
See Table~\ref{summary_of_design} for a summary of suggested designs, and Figure~\ref{fig:temi} for a condensed set of recommendations. 
\subsubsection{Robot vs. Environment}
The following privacy design themes focus on a robot’s ability to operate based on its physical environment and observed or programmed dynamics among users in a shared space. 

\paragraph{\textit{``[Robot], I'm good. You can go back home.''}: Preserving privacy at home.} 
\label{spatial_privacy}
All participants worried that a household robot’s mobility and ability to roam would compromise the privacy they expected in their homes. Although many participants explained that these concerns applied to almost any device that could store and transmit data, they worried that a robot that could move on its own was more likely to encounter moments not meant to be witnessed or recorded. In addition to concerns over access, some participants felt that the robot’s roaming presence alone could feel invasive. For instance, 14\_1 said, \textit{Like, last night I was watching some movies... I was like, ‘I want to be alone’... even [the robot] just being in the [same space]… it’s kind of weird when you want to be alone. }

\smallskip
\noindent\textit{\textbf{Limiting where and when a robot can operate.}}
To maintain a sense of privacy in their homes, participants suggested that a household robot should have \textbf{restricted access to certain areas or only enter them during specified times} (\textit{e.g.}, no one is home). Participants provided different ideas for limiting a robot’s access, from suggesting that a robot should only travel along \textbf{p}\textbf{redetermined pathways} to setting \textbf{specific ‘active’ periods} during which the robot could perform certain tasks (\textit{e.g.}, cleaning). 
However, participants also noted that simple spatial or temporal restrictions could be too simplistic, as the appropriateness of a robot’s presence often depended on its activity.
For example, when considering the robot's usage during nighttime hours, 8\_2 suggested that it could remain off in most instances, but be equipped with \textit{“an app where we could, like… disable [the robot’s routine], turn it on.”} 

While this design feature was suggested by almost all households, it did not address underlying concerns about a robot’s ability to detect and collect private data if it were parked or waiting just outside a restricted space. 
Additionally, some participants found it unsettling to adapt their routines to accommodate the robot’s pathways or schedules rather than the other way around; 4\_2 described this as feeling “held hostage” in their own home.

\smallskip
\noindent\textit{\textbf{Establishing Home Bases}}
 Several participants identified the need for the robot to have a ‘home base’ or designated areas where it can charge or travel to when not in use or when asked to leave a space.
 While the specific location depended on the household layout, participants often suggested communal areas such as living rooms or low-traffic spaces (\textit{e.g.}, corners) where the robot could remain hidden.
 However, some noted that even in less private areas, the robot's presence could still be uncomfortable, regardless of whether it was actively collecting data.  
 For example, both 9\_2 and 9\_C1 agreed that the living room, though considered a communal space for their family, was an undesirable home base, with 9\_2 explaining, \textit{"Well, my dad’s always there, so he wouldn’t like [the robot] in there, so, um, maybe the basement [for charging].” 
}

\begin{keytakeaways}
\small
\noindent \textbf{Design Recommendation: Restricted Access in Households}

\begin{itemize}[label=\textbullet, leftmargin=1em, nosep, topsep=2pt]
    \item Users should be able to customize where and when the robot is allowed to operate.
    \item When not performing a task, the robot should support a roaming mode that follows user-defined pathways.
    \item The robot should allow users to designate a “safe space” to which it can retreat, in addition to its charging station.
\end{itemize}
\end{keytakeaways}

\paragraph{\textit{“Hey buddy, what’d you hear today?”}: Privacy within and beyond the home.} 
\label{family}
The notion that a household robot might be used by multiple users raised concerns among participants about data management and sharing among active users, household members who did not use the robot, and guests. 
Even participants who expected to interact less with the robot than other household members acknowledged that they would still co-exist with it and, therefore, be subject to some level of data collection.
To this end, most households provide at least one idea for keeping specific users’ information, preferences, and conversations concealed from others, even if it sometimes means providing additional identifiable information (\textit{e.g.}, faces, fingerprints) to the robot to do so. 
While many participants denied having specific secrets to hide from household members, they emphasized that individual data privacy options helped, with participant 1\_2 describing it as a way to \textit{“keep peace between”} housemates.
Furthermore, this type of identifiable information could also be used to limit guests’ access to and interactions with their robot. While most participants did not mind if their guests wanted to try out the robot, they expressed concerns about guests gaining access to users’ data without permission, and felt responsible for any mishaps that might occur if a robot were to misinterpret a guest’s command or relay any private information a guest shared with the household.

\smallskip
\noindent\textit{\textbf{Maintaining privacy between users.}} 
A majority of participants felt that having user-specific profiles or settings on a household robot could support individual privacy within multi-user households. Each user could have a unique profile that contains any data collected about them or that they provide to the robot, and that would be inaccessible to other users without specific permissions. Participants also noted that having separate user profiles could provide one way for individuals to tailor the robot to their personal preferences \textit{(e.g.}, one user wanting a robot that announces its presence, and another wanting a silent robot). 
When thinking through how a household robot might recognize a specific user and adjust its settings accordingly, most participants suggested that the robot should be able to recognize individuals based on audio, visual, or biometric input. 
Importantly, some participants noted that any household robot using this information should also be able to make judgments about where to retain data or how to react if multiple users were present, with 12\_1 stating that the feature would mean \textit{“nothing if [the robots only] recognize you,”} and needed to also recognize\textit{ “the person right next to you.”} 

Participants exhibited varying degrees of comfort with the robot being able to profile them based on this information. Some participants wanted the robot to recognize only their voices, while others felt that facial recognition was fine as long as their voices were not retained. Some participants expressed little concern about this type of data being retained, while others became more open to a robot collecting and retaining it if it meant having control over their individual data. While Household 11 had qualms about the robot recognizing them by voice, 11\_2 eventually noted that this type of information was already being collected by devices they owned and used (\textit{e.g.}, smartphones, smart speakers): 
11\_1 stated, \textit{“So for that extent, I think the profiling [feature] is acceptable,” because “it sounds like Alexa already has that capability.”} 

About half of households considered how a household robot might notify users of private messages or reminders they did not want other users to see or hear. Some participants thought the robot’s \textbf{notifications should be as vague as possible}, thus only stating when it had a message for a specific user without explaining what the message was about until given further permission. Others liked the idea of having the robot \textbf{send any user-specific notifications via text or to a connected application on another device}. Household 12 suggested combining both notification methods, with 12\_1 suggesting that the robot \textit{“alerts you, like, ‘hey, you have a reminder for today’”} and then \textit{“maybe send, like, a message to your phone or something like that.” }
Notably, participants who preferred external notifications did not express concern about the robot accessing unrelated data on their devices. Instead, several considered phone-based control and notification to increase the robot’s utility, particularly when they were away from home or when they were paying attention to the robot. As 6\_1 explained: 
\textit{“We’re getting messages on the phone all the time, regardless of if we are in the conversation or not… rather than [the robot] saying something, sending a notification to either the phone or the watch- that, I’d notice.”}

\smallskip
\noindent\textit{\textbf{Maintaining privacy of and with guests}} 
About half of households thought it was important to restrict guests’ interactions with a household robot in their homes. For the most part, none of these households had issues with guests interacting with the robot, provided the guests had permission and were supervised. However, participants felt less certain about instances where a guest might interact with a robot alone, making it difficult for household members to monitor the information exchanged between the guest and the robot. Some participants worried that guests might access private data. 
Some participants also considered the consequences of guests sharing sensitive information or experiencing misunderstandings or misuse of the robot. Multiple participants wondered if they would be responsible for their robot's mistakes: \textit{“Can somebody say, ‘it’s your fault [that the robot failed to do a task], it’s your [robot’s] fault, it’s your family’s fault,’ ...but it was never our responsibility to begin with?”} (3\_1) 

    When asked how a robot might detect guest interactions, some participants suggested using voice, facial, or biometric recognition to distinguish household members from individuals without a profile. To this end, some participants mentioned feeling uncomfortable about the robot collecting data on guests without explicit permission, and suggested either unplugging the robot or requiring guests’ consent for it to operate. Household 8 thought their guests might enjoy interacting with a robot, but would introduce the pros and cons of doing so first: \textit{“... I would want to say, ‘we have a [robot]... if you would like to see how it works, we’re happy to bring it out,’... let them know what it can and cannot do... and them letting them decide...” } (8\_1)
    
Several participants proposed a \textbf{“guest mode”} that could be activated when visitors were present. While its exact behavior varied by household, some participants preferred the robot to remain inactive without explicit permission from a household member, whereas others wanted to define specific limits on what the robot could do or say around guests.
A few participants thought this feature would benefit the family as well. For instance, Household 15 wanted any robot in their home to notify them if a guest attempted to interact with it: \textit{“I feel like [the notification] would probably be, like, very similar to how people would react- or, like, you have the duo-factor [authentication]... like, ‘hey, someone is trying to log into your account. Is this you?’”} (15\_2)

\begin{keytakeaways}
\small
\noindent \textbf{Design Recommendation: Respect Household Boundaries}

\begin{itemize}[label=\textbullet, leftmargin=1em, nosep, topsep=2pt]
    \item When recognizing individual users, the robot should notify only the task-assigning user, with communication deliberately limited or obfuscated to protect privacy.
    \item Users should be able to create personal profiles tailored to their privacy preferences. Users should not be able to access other users' profiles or the data stored in them.
    \item When non-household members are present, the robot should support a “guest mode” that prevents (1) access to user data and (2) task assignment without explicit household permission.
\end{itemize}
\end{keytakeaways}

\subsubsection{Robot vs. Users}
The following privacy design themes focus on both how users might interact with the robot and how users prefer the robot to interact with them. 

\paragraph{\textit{“I don’t think I will ever be completely comfortable”}: User-initiated actions.}
\label{user_to_robot}
Participants recommended various designs that would allow them to maintain control over the robot’s actions out of caution and general distrust in technology manufacturers, even if it meant taking on additional responsibilities. 
    However, some participants acknowledged that having control was necessary, but that the extra responsibilities and need to monitor a robot rendered them unlikely to trust its presence in their home: \textit{You shouldn’t need to tell [the robot] over and over again, you know, ‘don’t collect this,’ or ‘close your ears’ ...it’s a little too much, and you have to do that so often, and you just feel disrupted.  }(10\_2)

\smallskip
\noindent\textit{\textbf{User-reliant training.}} 
Many participants provided privacy designs that would \textbf{require users to train and tweak their robots over an extended period of time}, including having the robot learn the household’s routines to understand when it should or should not be in specific spaces and familiarize itself with voices and faces in various settings. In many of these instances, participants assumed that a household robot would have inherent capabilities to receive and learn from collected data over time. 
When asked how much effort they were willing to invest in training a robot, participants’ responses varied widely. Some preferred minimal involvement, such as sharing critical information (\textit{e.g.}, flagging objects the robot should avoid) or reviewing and removing non-essential data from the robot’s training set. Others, however, were willing to take on a more active, structured training role over one to two months. 10\_2 stated, \textit{"I understand, you know, it’s gonna take time [for the robot to] pick on the cues, you know, maybe a month or so… I would like to see progress, obviously, during that time..."} Similarly, 1\_2 reasoned that tailoring the robot to their household’s preferences would take time, but should show signs of improvement during the training phase: \textit{“I think the first couple of months, there is likely to be a learning curve... but when it is up and running, I want it to be like... [clicks tongue and opens palms to signify ‘all good’] okay!” }

\smallskip
\noindent\textit{\textbf{User-initiated data collection and management.}} 
All households discussed design features that would allow users to determine when a robot could collect data and what data should be retained, while increasing users’ involvement in and management of the robot. In particular, households want the robot to offer users the \textbf{ability to review, manage, and delete data} before it is stored in a long-term database for training purposes or shared with other users or third parties. Some participants suggested a nightly routine, and others felt that a monthly maintenance was more feasible. 
Although hesitant to assume responsibility for what information a robot might retain, participant 4\_1 still felt that having the option to manage or delete stored data was important, and compared their perception of a robot overhearing to a friend: \textit{"A friend might forget, because humans are forgetful. [The robot] is not gonna forget... But, we’ve got the option to totally delete things... it would be totally gone. Friends’ memories aren’t necessarily gonna be deleted."}

    Although participants valued a robot’s ability to learn over time, most were uncomfortable with sharing their data with manufacturers or third parties without explicit permission.
    While some participants acknowledged that developers likely relied on user data to improve their products, they still had reservations about how developers might use the data. Most participants instead preferred \textbf{their data be stored in local storage or a cloud-based server that manufacturers did not have access to}. Local storage was perceived as the most secure option, though a few participants noted that users would be responsible for reviewing and deleting data periodically to ensure the robot did not exhaust all storage space. 

Households also wanted the robot to \textbf{only operate or record when given permission} or enact a \textbf{‘privacy mode’} where no data would be recorded or processed even if the robot continued to perform other functions. Even though 7\_2 could see some utility in having a household robot that could generate personalized recommendations using observational data, they wanted to be able to tell the robot when to do so, \textit{“versus it taking [data] by itself and doing it.”} Sometimes, participants wanted control over certain actions a robot could perform beyond recording. In response to the potential of a robot accessing confidential data on a computer screen, 5\_C2 suggested \textit{“it could have this thing where it can’t look at screens unless you may tell it to look at it.”} 

\smallskip
\noindent\textit{\textbf{Physical controls on the robot.}}
Despite proposing design solutions, participants did not believe that training and user-initiated commands alone were sufficient to ensure privacy-aware robot behavior. About half of the households suggested that the robot perform \textbf{physical actions to indicate when it was not recording or listening}, such as turning its camera away from a room or tilting its camera or other sensors toward the ceiling to show that it was not collecting any visual data. However, almost all households wanted \textbf{physical design features that would affect a robot’s ability to collect and retain data}, including built-in on/off buttons, camera lens covers, and removable data storage. Multiple people drew comparisons to how they managed smart speakers in their homes, pointing out that they did not truly believe that the devices were not listening unless they were unplugged. Furthermore, 8\_1 felt most assured by having some physical way to cut off the robot’s access to information: \textit{“As long as I could shut it off, I wouldn’t care [if it could move]... I’d have to be able to physically get to it to turn it back on then… I think an on-off button is always necessary.” }

\begin{keytakeaways}
\small
\noindent \textbf{Design Recommendation: Return Control to Users}

\begin{itemize}[label=\textbullet, leftmargin=1em, nosep, topsep=2pt]
    \item Users should be able to control data collection and management through training interactions and physical controls (\textit{e.g.}, buttons or covers) to pause sensing.
    \item The robot should store data locally and retain it only for task execution or learning, and must not share data with manufacturers without explicit user opt-in.
\end{itemize}
\end{keytakeaways}

\paragraph{\textit{“If you know what the rules of the game are, then you can play [the right] cards”}: Robot-initiated actions.} 
\label{robot_to_user}
Participants viewed a household robot that proactively notified users of its actions, policies, and presence, and found it not only more trustworthy but also easier to control and navigate. %
Many participants wanted final authority over what a robot could do in their homes, while several emphasized the need for developers to help users understand its capabilities to avoid unintended data collection. When faced with the question of how they might manage the robot if it were to do something wrong, 10\_1 said, \textit{“...It’s fun that [the robot] has its own autonomy... I’d let my dog do whatever she wants, and so that’s how I would treat this... if it’s endangering me or other people... we can talk about it. But... how?”} (10\_1) 

\smallskip
\noindent\textit{\textbf{Toward true transparency: accessibility and frequency.}}
Beyond the general desire for a robot to communicate or explain its behaviors and processes, several participants indicated that this alone was insufficient. Rather, they clarified that for a robot to be truly transparent, any written or verbal explanations or customization options needed to be easy to access and provided in plain language. 8\_1 
compared to their own experience reading terms of services: \textit{“... you give up a lot [of control] because you’ve got 19 pages of privacy policy to try to read through… if you were somebody that wanted to read the whole thing, do you understand 90\% of the jargon?” }

\smallskip
\noindent\textit{\textbf{Awareness around data collection and presence.}}
Most participants wanted the robot to \textbf{notify them when it was collecting or had collected data}. They emphasized the need for some form of notification when the robot was collecting audio or visual data. 3\_2 explained,\textit{"I get [a notification] that’s saying, ‘Hey, [the robot’s] recording,’ and I’m like, ‘Wait, I wasn’t wanting to be recorded.’"} Additionally, some participants wanted the robot to provide a clear signal when it was not collecting data, with suggestions ranging from showing a sleeping face or a moon image to indicate that it was ‘asleep’. 

	Participants differed on how they wanted to be notified, from visual indicators (\textit{e.g}., lights or symbols) to verbal announcements about something happened. While some suggestions, such as having an icon of a camera pop up to indicate data collection, were proposed more often, there was no perfect solution around how best to notify users. Whereas some participants disliked being interrupted by notifications and preferred something more subtle, others welcomed the discomfort and disruptions that came with more conspicuous indicators. Several households considered more intrusive notifications (\textit{e.g.}, asking for verbal permission before entering a room) more respectful of their privacy than simply assuming it was allowed to approach users or enter a shared space. 
    A few participants even noted that feeling uncomfortable around the robot helped them stay more mindful of their behavior around it. Sometimes, the idea of a robot having a face alone was enough of an unpleasant indicator that it was ‘conscious’ and capable of collecting data: \textit{“Maybe I kind of like [the face] because it reminds me that it is passively listening all the time… it’s like a reminder [that it’s on].”} (7\_1)

\begin{keytakeaways}
\small
\noindent \textbf{Design Recommendation: Transparency in Interaction}

\begin{itemize}[label=\textbullet, leftmargin=1em, nosep, topsep=2pt]
    \item Provide concise, accessible guidance explaining why data are collected, how collection occurs, and where the data go, with the option for users to review this information at any time.
    \item The robot should use clear visual, auditory, or behavioral cues (\textit{e.g.}, lights, icons, sounds, or gestures) to signal when collection begins and is ongoing, based on the user's preference.
\end{itemize}
\end{keytakeaways}

\section{Influential Mechanisms on Design Decisions}

Our findings reveal that, while most households managed to come up with privacy designs they would like to see implemented in a household robot, they would at times debate the efficacy and implications of those designs in practice. In this section, we discuss four mechanisms that moderate participants’ decisions about whether to implement certain types of privacy designs: 1) the value and privacy risks of owning a household robot, 2) the anthropomorphizing and emotional valuation of robots, 3) the functionality of a robot in a shared environment, and 4) expected shifts in perception and privacy concerns with long-term use and exposure.

\subsection{(In)convenience versus privacy}
Throughout the design process, participants would evaluate their privacy designs against the conveniences that a household robot might afford them from two perspectives: 1) if the benefits of a household robot would be worth giving up specific privacy measures, and 2) if the benefits of a household robot would be worth any extra labor or maintenance needed to maintain user privacy. 

Discussions of trade-offs between privacy and functionality occurred most frequently when participants began to imagine use cases for a household robot. Several households re-evaluated their initial resistance to having a robot enter their bathrooms or bedrooms, assuming that a household robot may one day be able to clean or organize their spaces for them. A couple of households that initially objected to recording capabilities later identified scenarios in which facial recognition could enhance household safety and privacy. 15\_1 suggested that a robot that could retain information about faces could be useful for locating household members: \textit{“If you have a bigger family and you’re, like, ‘find this person,’... and then all of the sudden, the robot can’t find [them]... Maybe... we need to be concerned.”} Interestingly, several participants also emphasized that they would prefer a more simple household robot with fewer data collection sensors, even if it meant the robot might be more functionally limited: \textit{“... the less features it possesses to identify stuff on its own, the more comfortable I’d be with having [the robot] in more spaces.” }(12\_1). By the end of sessions, some participants discussed how they might be more relaxed about certain privacy designs if they knew that the robot would add value to their lives, while others concluded that they could not think of any task a household robot could do for them that would be worth sacrificing their sense of privacy and security for. 

Most participants recognized that certain privacy designs they suggested would require users to take on more active roles managing or controlling an in-home robot, and asserted that they would prefer the additional labor over having less authority over the robot’s data collection, retention, and sharing processes. However, multiple participants stated that they would not consider owning a robot that required that much maintenance to begin with. 10\_1, for example, believed that purchasing a household robot would mean accepting its privacy risks, even if data management options were available: \textit{“Thinking of all of the stuff [the robot] could do... I don’t want to be in charge of that... I know what I’m getting into. Like, okay, [the company] can have [the data]. I don’t care.” }

To minimize user burden, some participants thought the robot should be able to automatically recognize or assess the privacy of a situation using nonverbal cues. For instance, the robot could stop recording whenever it noticed a user on the phone. Others, however, felt the robot should be able to engage in more sophisticated decision-making automation, such as being able to recognize social cues and users’ body language (\textit{e.g.}, leaning into someone’s ear to tell a secret) or make assumptions about a user’s well-being based on the tone of their voice. 
On this note, many participants who thought the robot could appraise situations on its own using these social cues or norms considered their own suggestions far-fetched and doubted that a household robot could successfully emulate human-like social judgments and behaviors.

\subsection{Social acceptability, empathy, and robots}
How participants thought the robot should be treated and addressed had both small-scale and wider implications for their privacy designs. First, almost all participants used niceties like ‘please’ and ‘thank you’ when giving an example of how they would request the robot to do something. In many cases, participants seemed unaware that they had done so until the facilitator asked why they had included these terms. Most people reasoned that they were used to using polite language when asking for a favor and thought it would be odd to change their behaviors just because they were interacting with a robot: \textit{“I think politeness is a habit… we say ‘please’ leave to the dog, too…”} (13\_1) A few joked that they did not think it would be wise to be rude to robots if they one day gained consciousness: \textit{“… I’d be nice to [the robots] because when they take over… I got to survive... everyone should be nice to [them]… just in case. [laughs]”}  (2\_1) 

Second, participants had mixed feelings about employing certain designs that could be perceived as impolite or cruel. Household 2 suggested several gestures (\textit{e.g.}, wave a hand to dismiss) before 2\_1 interjected that certain ones would feel belittling if done to another person. Similarly, a few households agreed they would not want any gestures at all, as they would not rely on them alone to communicate with someone else. Notably, Household 1 thought that gestures would be helpful, but commented on the fact that they would consider the robot ‘a tool’ and not anything more. 1\_2 pointed out that, if the robot were to exhibit more human-like qualities, they would likely have more mixed feelings: \textit{“There’s no consciousness… [but] I think it’s trying [to emulate it], because it does things that we usually do.”} A couple of other households also mentioned that they might struggle to enact certain privacy designs if they became attached to the robot, with 10\_1 saying, \textit{“I think I would feel bad turning it off… like, we both named our cars. Like, that’s an extension of us. To me, it’s like, car, pet, social robot, family. That’s my ranking.”} When 2\_2 was asked if they wanted the robot to only retain essential information to function, they said, \textit{“I don’t know. It’d make me sad that it can’t remember small things. I know it doesn’t have feelings, but it would just make me sad.”
}
\subsection{Navigating multi-user environments}
When designing a shared robot to meet all their privacy needs and preferences, households often engage in debates and negotiations over how the robot should be managed and controlled. In addition to disagreements over the specific design of a privacy feature (\textit{e.g.}, saving data for months versus years), participants sometimes failed to reach consensus on which rooms a robot should be able to enter or roam, and indicated different access preferences on their floor plan. Other participants were less concerned about specific areas of the house, but highlighted hypothetical instances where one user might take advantage of the robot’s mobility and sensors to monitor another user without their knowledge or consent: \textit{"How do you know if [another user] didn’t turn it on and it’s recording and he’s going to come home later and listen to what [roommate] did during the day or watch it?"} (8\_1)

Households also wondered how a robot would handle multiple users being present at once. 7\_1 felt that if a robot was going to exist in a shared environment, it should be able to determine what counted as a command and what did not: \textit{“[7\_C] just talks non-stop all the time, so I wouldn’t know how the robot is gonna discern who’s actually talking to it? When so many people are having overlapping conversations.”} Relatedly, Household 14 had qualms about how a robot would respond to conflicting commands from different users, and whether there should be a hierarchy of control among users. In response to 14\_1 suggesting that the robot could be sent to their children’s rooms to bug or monitor them, 14\_C countered that they would tell the robot to “turn off” to protect their privacy. This prompted 14\_1 to ponder which users could turn the robot on and off: \textit{"Because then I could just see us [14\_1 and 14\_C] going back and forth...''}

Among the eight households with at least one minor, six discussed parental controls tailored to their children’s ages and developmental levels. For younger children, participants favored more restrictive designs, such as disabling the robot’s interface or preventing conversational interaction. In households with older children or adolescents, parents preferred permission-based controls for sensitive tasks and the ability to review interactions. Participants noted that these preferences would likely evolve as their children grew up.


\subsection{Long-term exposure and experience} 
Most participants expected that their perceptions and privacy preferences for a household robot would shift with greater familiarity and time. Multiple people discussed how they had become more comfortable with the robot after limited interactions during the study, and felt that a longer in-home trial would help them better understand whether their privacy designs were appropriate to be adjusted to suit their needs. A couple of participants also mentioned that time was a necessary component for building any type of trust, and whether they might become more lenient with the robot’s privacy designs would depend on their experiences with the technology: \textit{"At this point, we don’t know the parameters of what’s being recorded and where it’s going, all that. Once we know that, then I think trustworthiness will improve."} (3\_1) 

Household 2 believed that, over time, they would become so accustomed to the robot that they would not notice if it was capturing private data or potentially sensitive information (e.g., financial documents). Meanwhile, the children in Household 4 thought that they would find the robot novel and fun at first, but would get used to its presence over time, with their parent, 4\_1, suggesting that they would \textit{“start ignoring it.”} A couple of households also assumed that the robot itself would learn new skills, which might be preferred if it could \textit{“really understand how to respect boundaries”} (9\_2), but perhaps less so if it began to exhibit more qualities that resembled human sentience: \textit{...Will it ask questions like ‘ What does it mean to live?’, ‘ How do I find happiness?’ [laughs] Like, do we have control of that? Is it going to do it on its own?} (3\_2)

\section{Discussion}
Our findings, drawn from the 15 households that completed the study, revealed that participants had distinct concerns about being able to control and maintain their data privacy around a robot that was shared across multiple users in the home (RQ1), and offered specific safety designs and risk management strategies that manufacturers could deploy in robots (RQ2). Importantly, participants not only discussed issues around manufacturer overreach and opaque data collection or sharing practices, but also potential misuse by different users in the home that might jeopardize their own or other users’ privacy. In working with households with limited or no experience with household robots, we also found that households expected their privacy perceptions of and preferences for the robot to change over time as knowledge, experience, and familiarity increased (RQ3). We note that some of our findings align well with those of Bell et al.~\cite{bell2025always} and extend their work to a multi-user approach. In particular, we observed households' discussions around spatial privacy and boundaries, with special regard to a household robot's mobility and user control of the robot. The following sections explore potential implications of the current work, design recommendations for household robotics developers, and methodological limitations that should be addressed in future research.

\subsection{Multi-user design perspectives}
By focusing on households in this study, we observed participants’ discussions of different shared-use scenarios for the robot, associated privacy concerns, and disagreements over preferred privacy designs, which would have been difficult to explore from a single perspective. We also noticed that participants would use shared knowledge about their households to contextualize their design suggestions (\textit{e.g.}, spatial limitations in homes, personal routines). While we found recurring themes related to privacy across our sample of households, each household unit provided ideas shaped by the specific identities and experiences present in its living space. 

	We also found numerous instances in which two or more members of the same household held different expectations about what a robot should do to maintain their privacy. While a household may be composed of individuals unified by a shared living space, each individual also exists beyond this space and brings their own social roles, identities, and experiences~\cite{stets2000identity}. Thus, it was unsurprising that we not only observed differences in how households wanted to approach certain privacy risks, but also that members in the same household held incongruent ideas and engaged in debates about how to handle a household robot.  
    
	Well-designed, privacy-aware robots for household use should account for the preferences and well-being of not only their primary user(s), but also non-users within and outside of the home. Although some of the privacy designs that participants highlighted would be most useful and relevant for users (\textit{e.g.}, having control over training the robot), many of the concerns that were floated could be applicable to anyone in proximity of the robot (\textit{e.g.}, having audio or visual data collected and shared without prior knowledge or consent). Indeed, prior research on smart speakers has highlighted instances in which one person in a household brought the device home for personal use, only to realize that others in the household would face the same privacy risks~\cite{meng2023multi}. Whereas smart speakers tend to stay in one place, many robots can relocate themselves and, as such, may have wider access to personal data around the household. Thus, manufacturers who intend to create robots for a broad consumer market must be conscientious about respecting the privacy of the spaces these devices will operate in.   

\subsection{Fluidity in privacy perceptions and design}
While household robots do exist, our participants found them relatively novel and discussed privacy design ideas, assuming their perceptions and preferences would evolve with greater familiarity with and knowledge of the technology. This allowed us to explore their initial reactions to a household robot. These interactions not only motivated them to discuss how their perceptions of the robot had changed over the course of the study, but also how they might change further if they had even more hands-on experience with the robot at home. Over time, participants imagined they might either learn about new privacy risks or become trusting enough of the robot that they would not mind reducing the level of control they wanted over the robot’s actions. Households continually hinted at the need for continual re-evaluations of their privacy designs, and hoped for devices that would allow such flexibility. 

	From our findings, we posit that flexible privacy design is characterized by a reasonable set of options and the ability to customize how these options are implemented throughout the device’s lifecycle. From our 15 study sessions, we generated an extensive list of privacy designs. Some of the most common designs were highlighted in our results, but even these varied in how participants discussed and thought about them. For instance, many households wanted the robot to be transparent about its policies and actions, but offered contrasting recommendations on how the robot should do so (\textit{e.g.}, announcing itself with sounds or words versus using iconography to minimize disruptions). To this end, it is unlikely that a single robot will be capable of providing every possible privacy design that a user could want. Furthermore, offering controls that are too open-ended or comprehensive can be overwhelming for users, and make them feel less confident and willing to engage in the customization process~\cite{korff2014too, chernev2015choice}. Rather than striving for a ‘perfect’ privacy-aware robot, developers should collaborate with potential consumers and non-users to develop proactive designs that address the most pertinent privacy concerns. 

\subsection{Limitations and Future Directions}
The current study was not without limitations. Although our sample size was comparable to that of other qualitative works on privacy designs for in-home technologies (\textit{e.g.}, ~\cite{yao2019defending, meng2021owning, dunbar2021someone}), our participants were predominantly White and well educated. Prior research suggests that an individual’s racial or ethnic identity and socioeconomic background may be associated with their privacy perceptions, preferences, and practices (\textit{e.g.}, ~\cite{wang2024online, guberek2018keeping}), as historically marginalized populations have often faced greater barriers to accessing educational resources or other supports. 
Future work on this topic may consider targeted recruitment of specific, underrepresented populations to support the field’s understanding of how identity may affect conceptions of in-home privacy around robots. 

    We note that while our co-design method supported a multi-user perspective on robot privacy design, it is inherently limited in generalizability. Future work may complement this approach with methods such as surveys to better capture broader population perspectives.
	Finally, our participants only interacted with the robot for a single study session in a lab setting. 
    Thus, participants had to speculate about long-term use.
    Participants also did not observe the robot in a familiar setting; although the lab resembled a living room, it did not reflect the unique layouts or conditions of their homes.
    Future longitudinal research investigating households’ perceptions of a privacy-aware robot in living spaces is planned.

\section{Conclusion}
As household robots become more common, it is increasingly imperative to evaluate the privacy risks they pose. By interacting with a household robot, participants in this study identified privacy concerns about the robot’s collection and use of personal data, and its ability to navigate and protect individual users’ privacy in a shared living space. Participants offered a wide range of strategies and designs to minimize privacy risks and discussed whether implementing these ideas would meet their needs and desires over time. We explore the implications of these findings, highlighting the diversity in perceptions and designs both between and within households, and offer design recommendations to uphold user privacy. 
\cleardoublepage
\bibliographystyle{plain}
\bibliography{main}

@inproceedings{bernd2020bystanders,
  title={$\{$Bystanders’$\}$ privacy: the perspectives of nannies on smart home surveillance},
  author={Bernd, Julia and Abu-Salma, Ruba and Frik, Alisa},
  booktitle={10th USENIX Workshop on Free and Open Communications on the Internet (FOCI 20)},
  year={2020}
}

@article{ahmad2020tangible,
  title={Tangible privacy: Towards user-centric sensor designs for bystander privacy},
  author={Ahmad, Imtiaz and Farzan, Rosta and Kapadia, Apu and Lee, Adam J},
  journal={Proceedings of the ACM on Human-Computer Interaction},
  volume={4},
  number={CSCW2},
  pages={1--28},
  year={2020},
  publisher={ACM New York, NY, USA}
}

@inproceedings{marky2020don,
  title={” I don’t know how to protect myself”: Understanding Privacy Perceptions Resulting from the Presence of Bystanders in Smart Environments},
  author={Marky, Karola and Voit, Alexandra and St{\"o}ver, Alina and Kunze, Kai and Schr{\"o}der, Svenja and M{\"u}hlh{\"a}user, Max},
  booktitle={Proceedings of the 11th Nordic Conference on Human-Computer Interaction: Shaping Experiences, Shaping Society},
  pages={1--11},
  year={2020}
}

@inproceedings{lee2016understanding,
  title={Understanding user privacy in Internet of Things environments},
  author={Lee, Hosub and Kobsa, Alfred},
  booktitle={2016 IEEE 3rd World Forum on Internet of Things (WF-IoT)},
  pages={407--412},
  year={2016},
  organization={IEEE}
}

@inproceedings{emami2019exploring,
  title={Exploring how privacy and security factor into IoT device purchase behavior},
  author={Emami-Naeini, Pardis and Dixon, Henry and Agarwal, Yuvraj and Cranor, Lorrie Faith},
  booktitle={Proceedings of the 2019 CHI Conference on Human Factors in Computing Systems},
  pages={1--12},
  year={2019}
}

@article{mare2020smart,
  title={Smart devices in Airbnbs: Considering privacy and security for both guests and hosts},
  author={Mare, Shrirang and Roesner, Franziska and Kohno, Tadayoshi},
  journal={Proceedings on Privacy Enhancing Technologies},
  year={2020}
}

@inproceedings{zeng2019understanding,
  title={Understanding and improving security and privacy in $\{$multi-user$\}$ smart homes: A design exploration and $\{$in-home$\}$ user study},
  author={Zeng, Eric and Roesner, Franziska},
  booktitle={28th USENIX Security Symposium (USENIX Security 19)},
  pages={159--176},
  year={2019}
}

@inproceedings{tabassum2019don,
  title={" I don't own the data": End User Perceptions of Smart Home Device Data Practices and Risks},
  author={Tabassum, Madiha and Kosinski, Tomasz and Lipford, Heather Richter},
  booktitle={Fifteenth symposium on usable privacy and security (SOUPS 2019)},
  pages={435--450},
  year={2019}
}

@inproceedings{yao2019defending,
  title={Defending my castle: A co-design study of privacy mechanisms for smart homes},
  author={Yao, Yaxing and Basdeo, Justin Reed and Kaushik, Smirity and Wang, Yang},
  booktitle={Proceedings of the 2019 chi conference on human factors in computing systems},
  pages={1--12},
  year={2019}
}

@article{abdi2019more,
  title={More than smart speakers: security and privacy perceptions of smart home personal assistants. Proceedings of the Fifteenth USENIX Conference on Usable Privacy and Security},
  author={Abdi, N and Ramokapane, KM and Such, JM},
  year={2019}
}

@article{saqib2025bystander,
  title={Bystander Privacy in Smart Homes: A Systematic Review of Concerns and Solutions},
  author={Saqib, Eimaan and He, Shijing and Choy, Junghyun and Abu-Salma, Ruba and Such, Jose and Bernd, Julia and Javed, Mobin},
  journal={ACM Transactions on Computer-Human Interaction},
  year={2025},
  publisher={ACM New York, NY}
}

@article{pattnaik2024security,
  title={Security and Privacy Perspectives of People Living in Shared Home Environments},
  author={Pattnaik, Nandita and Li, Shujun and Nurse, Jason RC},
  journal={Proceedings of the ACM on Human-Computer Interaction},
  volume={8},
  number={CSCW2},
  pages={1--39},
  year={2024},
  publisher={ACM New York, NY, USA}
}

@article{malkin2019privacy,
  title={Privacy attitudes of smart speaker users},
  author={Malkin, Nathan and Deatrick, Joe and Tong, Allen and Wijesekera, Primal and Egelman, Serge and Wagner, David},
  journal={Proceedings on Privacy Enhancing Technologies},
  volume={2019},
  number={4},
  year={2019}
}

@article{zheng2018user,
  title={User perceptions of smart home IoT privacy},
  author={Zheng, Serena and Apthorpe, Noah and Chetty, Marshini and Feamster, Nick},
  journal={Proceedings of the ACM on human-computer interaction},
  volume={2},
  number={CSCW},
  pages={1--20},
  year={2018},
  publisher={ACM New York, NY, USA}
}

@article{yao2019privacy,
  title={Privacy perceptions and designs of bystanders in smart homes},
  author={Yao, Yaxing and Basdeo, Justin Reed and Mcdonough, Oriana Rosata and Wang, Yang},
  journal={Proceedings of the ACM on Human-Computer Interaction},
  volume={3},
  number={CSCW},
  pages={1--24},
  year={2019},
  publisher={ACM New York, NY, USA}
}

@inproceedings{emami2023consumers,
  title={Are Consumers Willing to Pay for Security and Privacy of $\{$IoT$\}$ Devices?},
  author={Emami-Naeini, Pardis and Dheenadhayalan, Janarth and Agarwal, Yuvraj and Cranor, Lorrie Faith},
  booktitle={32nd USENIX Security Symposium (USENIX Security 23)},
  pages={1505--1522},
  year={2023}
}

@article{lau2018alexa,
  title={Alexa, are you listening? Privacy perceptions, concerns and privacy-seeking behaviors with smart speakers},
  author={Lau, Josephine and Zimmerman, Benjamin and Schaub, Florian},
  journal={Proceedings of the ACM on human-computer interaction},
  volume={2},
  number={CSCW},
  pages={1--31},
  year={2018},
  publisher={ACM New York, NY, USA}
}

@inproceedings{barbosa2020privacy,
  title={Do privacy and security matter to everyone? quantifying and clustering $\{$User-Centric$\}$ considerations about smart home device adoption},
  author={Barbosa, Nat{\~a} M and Zhang, Zhuohao and Wang, Yang},
  booktitle={Sixteenth Symposium on Usable Privacy and Security (SOUPS 2020)},
  pages={417--435},
  year={2020}
}

@inproceedings{zeng2017end,
  title={End user security and privacy concerns with smart homes},
  author={Zeng, Eric and Mare, Shrirang and Roesner, Franziska},
  booktitle={thirteenth symposium on usable privacy and security (SOUPS 2017)},
  pages={65--80},
  year={2017}
}

@inproceedings{li2023s,
  title={“It’s up to the Consumer to be Smart”: Understanding the Security and Privacy Attitudes of Smart Home Users on Reddit},
  author={Li, Jingjie and Sun, Kaiwen and Huff, Brittany Skye and Bierley, Anna Marie and Kim, Younghyun and Schaub, Florian and Fawaz, Kassem},
  booktitle={2023 IEEE Symposium on Security and Privacy (SP)},
  pages={2850--2866},
  year={2023},
  organization={IEEE}
}

@article{wang2025privacyguard,
  title={PrivacyGuard: Exploring Hidden Cross-App Privacy Leakage Threats In IoT Apps},
  author={Wang, Zhaohui and Luo, Bo and Li, Fengjun},
  journal={Proceedings on Privacy Enhancing Technologies},
  year={2025}
}

@article{aaraj2024vbit,
  title={VBIT: Towards Enhancing Privacy Control Over IoT Devices},
  author={Aaraj, Jad Al and Figueira, Olivia and Le, Tu and Figueira, Isabela and Trimananda, Rahmadi and Markopoulou, Athina},
  journal={arXiv preprint arXiv:2409.06233},
  year={2024}
}

@article{apthorpe2018discovering,
  title={Discovering smart home internet of things privacy norms using contextual integrity},
  author={Apthorpe, Noah and Shvartzshnaider, Yan and Mathur, Arunesh and Reisman, Dillon and Feamster, Nick},
  journal={Proceedings of the ACM on interactive, mobile, wearable and ubiquitous technologies},
  volume={2},
  number={2},
  pages={1--23},
  year={2018},
  publisher={ACM New York, NY, USA}
}

@article{sullivan2025benchmarking,
  title={Benchmarking LLM Privacy Recognition for Social Robot Decision Making},
  author={Sullivan, Dakota and Zhang, Shirley and Li, Jennica and Kirkorian, Heather and Mutlu, Bilge and Fawaz, Kassem},
  journal={arXiv preprint arXiv:2507.16124},
  year={2025}
}

@inproceedings{emami2020ask,
  title={Ask the experts: What should be on an IoT privacy and security label?},
  author={Emami-Naeini, Pardis and Agarwal, Yuvraj and Cranor, Lorrie Faith and Hibshi, Hanan},
  booktitle={2020 IEEE Symposium on Security and Privacy (SP)},
  pages={447--464},
  year={2020},
  organization={IEEE}
}

@inproceedings{marky2024decide,
  title={Decide yourself or delegate-user preferences regarding the autonomy of personal privacy assistants in private IoT-equipped environments},
  author={Marky, Karola and St{\"o}ver, Alina and Prange, Sarah and Bleck, Kira and Gerber, Paul and Zimmermann, Verena and M{\"u}ller, Florian and Alt, Florian and M{\"u}hlh{\"a}user, Max},
  booktitle={Proceedings of the 2024 CHI Conference on Human Factors in Computing Systems},
  pages={1--20},
  year={2024}
}

@inproceedings{albayaydh2024co,
  title={$\{$Co-Designing$\}$ a Mobile App for Bystander Privacy Protection in Jordanian Smart Homes: A Step Towards Addressing a Complex Privacy Landscape},
  author={Albayaydh, Wael and Flechais, Ivan},
  booktitle={33rd USENIX Security Symposium (USENIX Security 24)},
  pages={4963--4980},
  year={2024}
}

@inproceedings{levinson2024snitches,
  title={Snitches get unplugged: Adolescents' privacy concerns about robots in the home are relationally situated},
  author={Levinson, Leigh and Nippert-Eng, Christena and Gomez, Randy and Sabanovi{\'c}, Selma},
  booktitle={Proceedings of the 2024 ACM/IEEE international conference on human-robot interaction},
  pages={423--432},
  year={2024}
}

@inproceedings{han2005educational,
  title={The educational use of home robots for children},
  author={Han, Jeonghye and Jo, Miheon and Park, Sungju and Kim, Sungho},
  booktitle={ROMAN 2005. IEEE International Workshop on Robot and Human Interactive Communication, 2005.},
  pages={378--383},
  year={2005},
  organization={IEEE}
}

@inproceedings{levinson2024surveying,
  title={Surveying Adult Perceptions of Privacy and Attitudes towards Social Robots in the Home},
  author={Levinson, Leigh and Barrett, Tyler and Gomez, Randy and Sabanovi{\'c}, Selma},
  booktitle={Companion of the 2024 ACM/IEEE International Conference on Human-Robot Interaction},
  pages={664--668},
  year={2024}
}

@inproceedings{denning2009spotlight,
  title={A spotlight on security and privacy risks with future household robots: attacks and lessons},
  author={Denning, Tamara and Matuszek, Cynthia and Koscher, Karl and Smith, Joshua R and Kohno, Tadayoshi},
  booktitle={Proceedings of the 11th international conference on Ubiquitous computing},
  pages={105--114},
  year={2009}
}

@article{lutz2019privacy,
  title={The privacy implications of social robots: Scoping review and expert interviews},
  author={Lutz, Christoph and Sch{\"o}ttler, Maren and Hoffmann, Christian Pieter},
  journal={Mobile Media \& Communication},
  volume={7},
  number={3},
  pages={412--434},
  year={2019},
  publisher={SAGE Publications Sage UK: London, England}
}

@article{bacser2024yes,
  title={“Yes, It’s Cute, But How Can I Be Sure It’s Safe or Not?” Investigating the Intention to Use Service Robots in the Context of Privacy Calculus},
  author={Ba{\c{s}}er, Mira{\c{c}} Y{\"u}cel and B{\"u}y{\"u}kbe{\c{s}}e, Tuba and Durmaz, Yakup},
  journal={International Journal of Human--Computer Interaction},
  volume={40},
  number={20},
  pages={6151--6166},
  year={2024},
  publisher={Taylor \& Francis}
}

@inproceedings{collins2024socially,
  title={“Socially Assistive Robot Privacy Model”: A Multi-model Approach to Evaluating Socially Assistive Robot Privacy Concerns},
  author={Collins, Sawyer and Stanojevi{\'c}, {\v{C}}edomir and Bennett, Casey and Henkel, Zachary and Henkel, Kenna Baugus and Abbott, Nikki M and Bethel, Cindy L and {\'S}abanovi{\'c}, Selma},
  booktitle={International Conference on Social Robotics},
  pages={280--289},
  year={2024},
  organization={Springer}
}

@article{randall2019survey,
  title={A survey of robot-assisted language learning (RALL)},
  author={Randall, Natasha},
  journal={ACM Transactions on Human-Robot Interaction (THRI)},
  volume={9},
  number={1},
  pages={1--36},
  year={2019},
  publisher={ACM New York, NY, USA}
}

@inproceedings{tang2022confidant,
  title={Confidant: A privacy controller for social robots},
  author={Tang, Brian and Sullivan, Dakota and Cagiltay, Bengisu and Chandrasekaran, Varun and Fawaz, Kassem and Mutlu, Bilge},
  booktitle={2022 17th ACM/IEEE International Conference on Human-Robot Interaction (HRI)},
  pages={205--214},
  year={2022},
  organization={IEEE}
}

@inproceedings{kennedy2015robot,
  title={The robot who tried too hard: Social behaviour of a robot tutor can negatively affect child learning},
  author={Kennedy, James and Baxter, Paul and Belpaeme, Tony},
  booktitle={Proceedings of the tenth annual ACM/IEEE international conference on human-robot interaction},
  pages={67--74},
  year={2015}
}

@inproceedings{kennedy2016social,
  title={Social robot tutoring for child second language learning},
  author={Kennedy, James and Baxter, Paul and Senft, Emmanuel and Belpaeme, Tony},
  booktitle={2016 11th ACM/IEEE international conference on human-robot interaction (HRI)},
  pages={231--238},
  year={2016},
  organization={IEEE}
}

@article{wada2007living,
  title={Living with seal robots—its sociopsychological and physiological influences on the elderly at a care house},
  author={Wada, Kazuyoshi and Shibata, Takanori},
  journal={IEEE transactions on robotics},
  volume={23},
  number={5},
  pages={972--980},
  year={2007},
  publisher={IEEE}
}

@article{shibata2011robot,
  title={Robot therapy: a new approach for mental healthcare of the elderly--a mini-review},
  author={Shibata, Takanori and Wada, Kazuyoshi},
  journal={Gerontology},
  volume={57},
  number={4},
  pages={378--386},
  year={2011},
  publisher={S. Karger AG Basel, Switzerland}
}

@article{moyle2018care,
  title={Care staff perceptions of a social robot called Paro and a look-alike Plush Toy: a descriptive qualitative approach},
  author={Moyle, Wendy and Bramble, Marguerite and Jones, Cindy and Murfield, Jenny},
  journal={Aging \& mental health},
  volume={22},
  number={3},
  pages={330--335},
  year={2018},
  publisher={Taylor \& Francis}
}

@article{belpaeme2018social,
  title={Social robots for education: A review},
  author={Belpaeme, Tony and Kennedy, James and Ramachandran, Aditi and Scassellati, Brian and Tanaka, Fumihide},
  journal={Science robotics},
  volume={3},
  number={21},
  pages={eaat5954},
  year={2018},
  publisher={American Association for the Advancement of Science}
}

@inproceedings{carros2020exploring,
  title={Exploring human-robot interaction with the elderly: results from a ten-week case study in a care home},
  author={Carros, Felix and Meurer, Johanna and L{\"o}ffler, Diana and Unbehaun, David and Matthies, Sarah and Koch, Inga and Wieching, Rainer and Randall, Dave and Hassenzahl, Marc and Wulf, Volker},
  booktitle={Proceedings of the 2020 CHI conference on human factors in computing systems},
  pages={1--12},
  year={2020}
}

@article{guerrero2017cybersecurity,
  title={Cybersecurity of robotics and autonomous systems: Privacy and safety},
  author={Guerrero, {\'A}ngel Manuel},
  journal={Robotics: legal, ethical and socioeconomic impacts},
  pages={75},
  year={2017},
  publisher={BoD--Books on Demand}
}

@article{chatzimichali2020toward,
  title={Toward privacy-sensitive human--robot interaction: Privacy terms and human--data interaction in the personal robot era},
  author={Chatzimichali, Anna and Harrison, Ross and Chrysostomou, Dimitrios},
  journal={Paladyn, Journal of Behavioral Robotics},
  volume={12},
  number={1},
  pages={160--174},
  year={2020},
  publisher={De Gruyter}
}

@article{bell2025always,
  title={" Is it always watching? Is it always listening?" Exploring Contextual Privacy and Security Concerns Toward Domestic Social Robots},
  author={Bell, Henry and Kwesi, Jabari and Laabadli, Hiba and Emami-Naeini, Pardis},
  journal={arXiv preprint arXiv:2507.10786},
  year={2025}
}

@article{pagallo2018rise,
  title={The rise of robotics \& AI: technological advances \& normative dilemmas},
  author={Pagallo, Ugo and Corrales, Marcelo and Fenwick, Mark and Forg{\'o}, Nikolaus},
  journal={Robotics, AI and the Future of Law},
  pages={1--13},
  year={2018},
  publisher={Springer}
}

@inproceedings{rueben2018themes,
  title={Themes and research directions in privacy-sensitive robotics},
  author={Rueben, Matthew and Aroyo, Alexander Mois and Lutz, Christoph and Schm{\"o}lz, Johannes and Van Cleynenbreugel, Pieter and Corti, Andrea and Agrawal, Siddharth and Smart, William D},
  booktitle={2018 IEEE workshop on advanced robotics and its social impacts (ARSO)},
  pages={77--84},
  year={2018},
  organization={IEEE}
}

@inproceedings{zitkovich2023rt,
  title={Rt-2: Vision-language-action models transfer web knowledge to robotic control},
  author={Zitkovich, Brianna and Yu, Tianhe and Xu, Sichun and Xu, Peng and Xiao, Ted and Xia, Fei and Wu, Jialin and Wohlhart, Paul and Welker, Stefan and Wahid, Ayzaan and others},
  booktitle={Conference on Robot Learning},
  pages={2165--2183},
  year={2023},
  organization={PMLR}
}

@article{driess2023palm,
  title={Palm-e: An embodied multimodal language model},
  author={Driess, Danny and Xia, Fei and Sajjadi, Mehdi SM and Lynch, Corey and Chowdhery, Aakanksha and Wahid, Ayzaan and Tompson, Jonathan and Vuong, Quan and Yu, Tianhe and Huang, Wenlong and others},
  year={2023}
}

@article{singh2022progprompt,
  title={Progprompt: Generating situated robot task plans using large language models},
  author={Singh, Ishika and Blukis, Valts and Mousavian, Arsalan and Goyal, Ankit and Xu, Danfei and Tremblay, Jonathan and Fox, Dieter and Thomason, Jesse and Garg, Animesh},
  journal={arXiv preprint arXiv:2209.11302},
  year={2022}
}

@article{wake2023chatgpt,
  title={Chatgpt empowered long-step robot control in various environments: A case application},
  author={Wake, Naoki and Kanehira, Atsushi and Sasabuchi, Kazuhiro and Takamatsu, Jun and Ikeuchi, Katsushi},
  journal={IEEE Access},
  volume={11},
  pages={95060--95078},
  year={2023},
  publisher={IEEE}
}

@inproceedings{carlini2021extracting,
  title={Extracting training data from large language models},
  author={Carlini, Nicholas and Tramer, Florian and Wallace, Eric and Jagielski, Matthew and Herbert-Voss, Ariel and Lee, Katherine and Roberts, Adam and Brown, Tom and Song, Dawn and Erlingsson, Ulfar and others},
  booktitle={30th USENIX security symposium (USENIX Security 21)},
  pages={2633--2650},
  year={2021}
}

@article{staab2023beyond,
  title={Beyond memorization: Violating privacy via inference with large language models},
  author={Staab, Robin and Vero, Mark and Balunovi{\'c}, Mislav and Vechev, Martin},
  journal={arXiv preprint arXiv:2310.07298},
  year={2023}
}

@article{huang2025survey,
  title={A survey on hallucination in large language models: Principles, taxonomy, challenges, and open questions},
  author={Huang, Lei and Yu, Weijiang and Ma, Weitao and Zhong, Weihong and Feng, Zhangyin and Wang, Haotian and Chen, Qianglong and Peng, Weihua and Feng, Xiaocheng and Qin, Bing and others},
  journal={ACM Transactions on Information Systems},
  volume={43},
  number={2},
  pages={1--55},
  year={2025},
  publisher={ACM New York, NY}
}

@misc{robotemi_temi_v3_2026,
  title        = {Introducing temi robot V3},
  author       = {{Robotemi}},
  howpublished = {\url{https://www.robotemi.com/product/temi-sales-contact/?srsltid=AfmBOopD5KDlwewsk_90LdsYDU-OoJMfrkaKYUvYGvw-n5Z9WYr7v0HK}},
  year         = {2026},
  note         = {Accessed: 2026-01-27},
  organization = {Robotemi},
}

@inproceedings{dahlback1993wizard,
  title={Wizard of Oz studies: why and how},
  author={Dahlb{\"a}ck, Nils and J{\"o}nsson, Arne and Ahrenberg, Lars},
  booktitle={Proceedings of the 1st international conference on Intelligent user interfaces},
  pages={193--200},
  year={1993}
}

@article{braun2006using,
  title={Using thematic analysis in psychology},
  author={Braun, Virginia and Clarke, Victoria},
  journal={Qualitative research in psychology},
  volume={3},
  number={2},
  pages={77--101},
  year={2006},
  publisher={Taylor \& Francis}
}

@misc{kaixxx_noScribe_2026,
  title        = {{noScribe}: Cutting Edge AI Technology for Automated Audio Transcription},
  author       = {{kaixxx}},
  howpublished = {\url{https://github.com/kaixxx/noScribe}},
  year         = {2026},
  note         = {GitHub repository, accessed 2026-01-27},
}

@misc{1xNeo2025,
  author       = {{1X Technologies}},
  title        = {{NEO Home Robot}},
  howpublished = {\url{https://www.1x.tech/neo}},
  note         = {Accessed: February 3, 2026},
  year         = {2025}
}

@misc{figure03News2025,
  author       = {{Figure AI}},
  title        = {{Introducing Figure 03}},
  howpublished = {\url{https://www.figure.ai/news/introducing-figure-03}},
  note         = {Accessed: February 3, 2026},
  year         = {2025}
}

@article{bolton2021security,
  title={On the security and privacy challenges of virtual assistants},
  author={Bolton, Tom and Dargahi, Tooska and Belguith, Sana and Al-Rakhami, Mabrook S and Sodhro, Ali Hassan},
  journal={Sensors},
  volume={21},
  number={7},
  pages={2312},
  year={2021},
  publisher={MDPI}
}

@inproceedings{seymour2023legal,
  title={Legal obligation and ethical best practice: Towards meaningful verbal consent for voice assistants},
  author={Seymour, William and Cote, Mark and Such, Jose},
  booktitle={Proceedings of the 2023 CHI Conference on Human Factors in Computing Systems},
  pages={1--16},
  year={2023}
}

@inproceedings{geeng2019s,
  title={Who's in control? Interactions in multi-user smart homes},
  author={Geeng, Christine and Roesner, Franziska},
  booktitle={Proceedings of the 2019 CHI conference on human factors in computing systems},
  pages={1--13},
  year={2019}
}

@inproceedings{windl2024privacy,
  title={Privacy communication patterns for domestic robots},
  author={Windl, Maximiliane and Leusmann, Jan and Schmidt, Albrecht and Feger, Sebastian S and Mayer, Sven},
  booktitle={Twentieth Symposium on Usable Privacy and Security (SOUPS 2024)},
  pages={121--138},
  year={2024}
}

@article{lutz2021privacy,
  title={Do privacy concerns about social robots affect use intentions? Evidence from an experimental vignette study},
  author={Lutz, Christoph and Tam{\`o}-Larrieux, Aurelia},
  journal={Frontiers in Robotics and AI},
  volume={8},
  pages={627958},
  year={2021},
  publisher={Frontiers Media SA}
}

@article{hu2025information,
  title={Information transparency, privacy concerns, and customers' behavioral intentions regarding AI-powered hospitality robots: A situational awareness perspective},
  author={Hu, Yaou and Min, Hyounae Kelly},
  journal={Journal of Hospitality and Tourism Management},
  volume={63},
  pages={177--184},
  year={2025},
  publisher={Elsevier}
}

@article{khezresmaeilzadeh2024echoes,
  title={Echoes of privacy: Uncovering the profiling practices of voice assistants},
  author={Khezresmaeilzadeh, Tina and Zhu, Elaine and Grieco, Kiersten and Dubois, Daniel J and Psounis, Konstantinos and Choffnes, David},
  journal={arXiv preprint arXiv:2409.07444},
  year={2024}
}

@article{acosta2022survey,
  title={A survey on privacy issues and solutions for Voice-controlled Digital Assistants},
  author={Acosta, Luca Hern{\'a}ndez and Reinhardt, Delphine},
  journal={Pervasive and Mobile Computing},
  volume={80},
  pages={101523},
  year={2022},
  publisher={Elsevier}
}

@inproceedings{le2024towards,
  title={Towards real-time voice interaction data collection monitoring and ambient light privacy notification for voice-controlled services},
  author={Le, Tu and Wang, Zixin and Huang, Danny and Yao, Yaxing and Tian, Yuan},
  year={2024},
  organization={Symposium on Usable Security and Privacy (USEC) 2024}
}

@inproceedings{hudig2025rights,
  title={Rights Out of Sight: Data Practices and Transparency Gaps in Smart Consumer IoT Ecosystems},
  author={Hudig, Anna Ida and Mandalari, Anna Maria and Norval, Chris and Haddadi, Hamed and Binns, Reuben and Singh, Jatinder},
  booktitle={Proceedings of the 2025 ACM Conference on Fairness, Accountability, and Transparency},
  pages={2260--2273},
  year={2025}
}

@misc{horowitzSmartDevices2025,
  author       = {{Horowitz Research}},
  title        = {{Nearly Half of American Homes Have Smart Devices, Higher Among Younger and Multicultural Consumers, New Horowitz Study Finds}},
  howpublished = {\url{https://www.horowitzresearch.com/all/nearly-half-of-american-homes-have-smart-devices-higher-among-younger-and-multicultural-consumers/}},
  note         = {Published May 13, 2025; Accessed: February 3, 2026},
  year         = {2025}
}

@inproceedings{park2023nobody,
  title={" Nobody's Happy": Design Insights from $\{$Privacy-Conscious$\}$ Smart Home Power Users on Enhancing Data Transparency, Visibility, and Control},
  author={Park, Sunyup and Lenhart, Anna and Zimmer, Michael and Vitak, Jessica},
  booktitle={Nineteenth Symposium on Usable Privacy and Security (SOUPS 2023)},
  year={2023}
}

@article{alshehri2022exploring,
  title={Exploring the privacy concerns of bystanders in smart homes from the perspectives of both owners and bystanders},
  author={Alshehri, Ahmed and Spielman, Joseph and Prasad, Amiya and Yue, Chuan},
  journal={Proceedings on Privacy Enhancing Technologies},
  year={2022}
}

@article{lenhart2023you,
  title={" You Shouldn't Need to Share Your Data": Perceived Privacy Risks and Mitigation Strategies Among Privacy-Conscious Smart Home Power Users},
  author={Lenhart, Anna and Park, Sunyup and Zimmer, Michael and Vitak, Jessica},
  journal={Proceedings of the ACM on Human-Computer Interaction},
  volume={7},
  number={CSCW2},
  pages={1--34},
  year={2023},
  publisher={ACM New York, NY, USA}
}

@article{shi2024large,
  title={Large language model safety: A holistic survey},
  author={Shi, Dan and Shen, Tianhao and Huang, Yufei and Li, Zhigen and Leng, Yongqi and Jin, Renren and Liu, Chuang and Wu, Xinwei and Guo, Zishan and Yu, Linhao and others},
  journal={arXiv preprint arXiv:2412.17686},
  year={2024}
}

@article{stapels2023never,
  title={Never trust anything that can think for itself, if you can’t control its privacy settings: The influence of a robot’s privacy settings on users’ attitudes and willingness to self-disclose},
  author={Stapels, Julia G and Penner, Angelika and Diekmann, Niels and Eyssel, Friederike},
  journal={International Journal of Social Robotics},
  volume={15},
  number={9},
  pages={1487--1505},
  year={2023},
  publisher={Springer}
}

@misc{futurismRobotVacuum2025,
  author       = {Joe Wilkins},
  title        = {{Man Alarmed to Discover His Smart Vacuum Was Broadcasting a Secret Map of His House}},
  howpublished = {\url{https://futurism.com/robots-and-machines/robot-vacuum-broadcasting}},
  note         = {Published Oct. 26, 2025; Accessed: February 3, 2026},
  year         = {2025}
}

@inproceedings{ostrowski2021long,
  title={Long-term co-design guidelines: empowering older adults as co-designers of social robots},
  author={Ostrowski, Anastasia K and Breazeal, Cynthia and Park, Hae Won},
  booktitle={2021 30th IEEE international conference on robot \& human interactive communication (RO-MAN)},
  pages={1165--1172},
  year={2021},
  organization={IEEE}
}

@misc{techreviewRoomba2022,
  author       = {Eileen Guo},
  title        = {{A Roomba recorded a woman on the toilet. How did screenshots end up on Facebook?}},
  howpublished = {\url{https://www.technologyreview.com/2022/12/19/1065306/roomba-irobot-robot-vacuums-artificial-intelligence-training-data-privacy/}},
  note         = {Published December 19, 2022; Accessed: February 3, 2026},
  year         = {2022}
}

@inproceedings{stegner2023situated,
  title={Situated participatory design: A method for in situ design of robotic interaction with older adults},
  author={Stegner, Laura and Senft, Emmanuel and Mutlu, Bilge},
  booktitle={Proceedings of the 2023 CHI Conference on Human Factors in Computing Systems},
  pages={1--15},
  year={2023}
}

@inproceedings{korff2014too,
  title={Too Much Choice:$\{$End-User$\}$ Privacy Decisions in the Context of Choice Proliferation},
  author={Korff, Stefan and B{\"o}hme, Rainer},
  booktitle={10th Symposium On Usable Privacy and Security (SOUPS 2014)},
  pages={69--87},
  year={2014}
}

@article{chernev2015choice,
  title={Choice overload: A conceptual review and meta-analysis},
  author={Chernev, Alexander and B{\"o}ckenholt, Ulf and Goodman, Joseph},
  journal={Journal of Consumer Psychology},
  volume={25},
  number={2},
  pages={333--358},
  year={2015},
  publisher={Elsevier}
}

@inproceedings{meng2023multi,
  title={Multi-User Smart Speakers-A Narrative Review of Concerns and Problematic Interactions},
  author={Meng-Schneider, Nicole and Yasa Kostas, Rabia and Vaniea, Kami and Wolters, Maria K},
  booktitle={Extended Abstracts of the 2023 CHI Conference on Human Factors in Computing Systems},
  pages={1--7},
  year={2023}
}

@inproceedings{antony2023co,
  title={Co-designing with older adults, for older adults: robots to promote physical activity},
  author={Antony, Victor Nikhil and Cho, Sue Min and Huang, Chien-Ming},
  booktitle={Proceedings of the 2023 ACM/IEEE International Conference on Human-Robot Interaction},
  pages={506--515},
  year={2023}
}

@inproceedings{obaid2023collective,
  title={Collective co-design activities with children for designing classroom robots},
  author={Obaid, Mohammad and Baykal, G{\"o}k{\c{c}}e Elif and K{\i}rlang{\i}c, G{\"u}ncel and G{\"o}ksun, Tilbe and Yanta{\c{c}}, As{\i}m Evren},
  booktitle={Proceedings of the 4th African Human Computer Interaction Conference},
  pages={229--237},
  year={2023}
}

@article{claudio2017interaction,
  title={Interaction design for cultural heritage. A robotic cultural game for visiting the museum’s inaccessible areas.},
  author={Claudio, Germak and Luca, Giuliano and Luce, Lupetti Maria},
  journal={The Design Journal},
  volume={20},
  number={sup1},
  pages={S3925--S3934},
  year={2017},
  publisher={Taylor \& Francis}
}

@article{stets2000identity,
  title={Identity theory and social identity theory},
  author={Stets, Jan E and Burke, Peter J},
  journal={Social psychology quarterly},
  pages={224--237},
  year={2000},
  publisher={JSTOR}
}

@article{meng2021owning,
  title={Owning and sharing: Privacy perceptions of smart speaker users},
  author={Meng, Nicole and Kek{\"u}ll{\"u}o{\u{g}}lu, Dilara and Vaniea, Kami},
  journal={Proceedings of the ACM on Human-Computer Interaction},
  volume={5},
  number={CSCW1},
  pages={1--29},
  year={2021},
  publisher={ACM New York, NY, USA}
}

@article{dunbar2021someone,
  title={Is someone listening? audio-related privacy perceptions and design recommendations from guardians, pragmatists, and cynics},
  author={Dunbar, Julia C and Bascom, Emily and Boone, Ashley and Hiniker, Alexis},
  journal={Proceedings of the ACM on Interactive, Mobile, Wearable and Ubiquitous Technologies},
  volume={5},
  number={3},
  pages={1--23},
  year={2021},
  publisher={ACM New York, NY, USA}
}

@article{wang2024online,
  title={The online privacy divide: testing resource and identity explanations for racial/ethnic differences in privacy concerns and privacy management behaviors on social media},
  author={Wang, Laurent H and Metzger, Miriam J},
  journal={Communication Research},
  pages={00936502241273157},
  year={2024},
  publisher={SAGE Publications Sage CA: Los Angeles, CA}
}

@inproceedings{guberek2018keeping,
  title={Keeping a low profile? Technology, risk and privacy among undocumented immigrants},
  author={Guberek, Tamy and McDonald, Allison and Simioni, Sylvia and Mhaidli, Abraham H and Toyama, Kentaro and Schaub, Florian},
  booktitle={Proceedings of the 2018 CHI conference on human factors in computing systems},
  pages={1--15},
  year={2018}
}

\end{document}